\documentclass[journal]{IEEEtran}

\usepackage[T1]{fontenc}
\usepackage[utf8]{inputenc}

\usepackage{amsmath}
\usepackage{amssymb}
\usepackage{amsfonts}

\usepackage{graphicx}
\usepackage{booktabs}
\usepackage{array}
\usepackage{tabularx}
\newcolumntype{R}{>{\raggedleft\arraybackslash}X}
\usepackage{multirow}
\usepackage{float}
\usepackage{enumitem}

\usepackage{microtype}
\usepackage{textcomp}
\usepackage{xcolor}
\usepackage{url}  

\usepackage{siunitx}
\usepackage{placeins}   

\usepackage[nospace]{cite}

\usepackage[colorlinks=false,pdfborder={0 0 0}]{hyperref}

\begin{document}

\renewcommand{\topfraction}{0.90}
\renewcommand{\bottomfraction}{0.50}
\renewcommand{\textfraction}{0.08}
\renewcommand{\floatpagefraction}{0.85}
\setcounter{topnumber}{2}
\setcounter{bottomnumber}{1}
\setcounter{totalnumber}{3}

\title{A Frequency-Controlled Comparison of Tick- and Minute-Based Information Bars for Cryptocurrency Markets}

\author{Muhammad~Toheed~Fayyaz, Abdul~Jabbar, ~Faheem~Ahmad~Qureshi, and~Syed~Qaisar~Jalil}

\maketitle

\begin{abstract}
Calendar-based bar construction maps a heterogeneous stream of financial
market activity onto a homogeneous temporal grid, introducing well-documented
statistical distortions that impede econometric analysis and machine learning
applications. Information-driven bars address these distortions by closing a
new bar whenever accumulated market activity reaches an adaptive threshold,
aligning the sampling clock with the underlying information-arrival process.
Despite widespread adoption of this framework, the effect of input data
resolution on bar quality has not been systematically studied: practitioners
routinely construct information bars from pre-aggregated one-minute OHLCV data
rather than from raw tick records, yet the two data sources produce
fundamentally different accumulation signals.
This paper provides a controlled comparison of six information bar
types (dollar, volume, volatility, range, Renko, and hybrid bars) constructed
from both raw Binance aggTrade tick data and one-minute OHLCV bars for the
BTCUSDT USDT-margined perpetual futures market over a six-year period spanning
January 2020 to December 2025, and evaluated against fixed-interval time-bar baselines.
Both pipelines share a common adaptive EMA calibration framework; the tick
pipeline additionally uses strictly tick-native activity signals, isolating
data resolution as the sole experimental variable.
Results across eight statistical quality criteria reveal that the tick
advantage is bar-type-specific and most pronounced in bar types whose activity
signals are most sensitive to intra-minute price dynamics: tick Renko bars
achieve the smallest random-walk deviation recorded
($|\mathrm{VR}(4){-}1| = 0.020$, lag-1 autocorrelation $= 0.002$), and tick
volatility bars reduce serial dependence by 69\% relative to the minute
baseline ($|\mathrm{VR}(4){-}1|: 0.028$ versus $0.089$).
In the multi-regime six-year sample, normality improvements are
regime-dependent and secondary: the extreme market events of 2020--2022
inflate fat tails across all bar types, and Ljung-Box independence is
rejected for all series at the sample sizes studied.
A matched-frequency robustness analysis shows that the apparent tick
underperformance on distributional criteria is largely a sampling-frequency
artefact: when tick series are coarsened to the minute pipeline's bar count,
frequency-matched tick dollar bars lead on all six criteria and matched tick
volatility bars attain LB $p = 0.51$, recovering serial independence that
the raw oversampled series rejects.
Calendar-bar baselines are a competitive reference set; information bars
deliver their primary advantage through superior variance-ratio conformance
and near-zero timeout rates, particularly for the tick pipeline.
The findings provide bar-type-specific guidance for practitioners choosing
between tick and minute data sources, with direct implications for
machine learning applications in cryptocurrency markets.
\end{abstract}



\section{Introduction}
\label{sec:introduction}

\IEEEPARstart{T}{he} representation of financial market data is not a
neutral act.
How a practitioner or researcher partitions the continuous stream of
trades and quotes into discrete observations fundamentally shapes the
statistical properties of the resulting series, the signals that any
downstream model can detect, and ultimately the inferences drawn from
the data.
Yet the dominant convention in both academic research and industry
practice remains the calendar-based bar: a fixed-interval
aggregation (most commonly daily, hourly, or one-minute) that maps a
heterogeneous flow of market activity onto a homogeneous temporal grid.
This choice, while administratively convenient, introduces a
well-documented set of distortions that are inconsistent with the
theoretical premises of modern market microstructure theory
\cite{OHara1995,Easley_OHara1992}.

Cryptocurrency markets make this problem especially acute.
Bitcoin~\cite{Nakamoto2008} and related assets trade continuously,
twenty-four hours a day, seven days a week, without the market-open
and market-close mechanisms that impose natural periodicity on equity
markets~\cite{Baur_etal2018}.
Order flow is highly uneven: quiet intervals of thin activity are
punctuated by sharp, news-driven bursts in which hundreds of trades
execute within seconds \cite{Baur_etal2018,Urquhart2016}.
Empirical studies of Bitcoin consistently document the stylised facts
that motivate departure from calendar sampling: fat-tailed return
distributions, long-range autocorrelation in absolute returns, and
persistent volatility clustering
\cite{Bariviera2017,Katsiampa2017,Urquhart2016}.
Partitioning such a process into fixed one-minute windows forces the
analyst to treat an interval of five hundred high-velocity trades
during a volatility spike identically to an interval of two passive
fills during a quiet overnight session, a conflation that is
theoretically unjustifiable and statistically damaging.

The statistical consequences of calendar sampling are well established
in the foundational literature.
Time-sampled returns exhibit serial correlation in absolute returns,
heteroskedasticity driven by the clustering of informative events, and
fat tails that persist even after volatility normalisation
\cite{Mandelbrot1963,Clark1973,Engle1982,Bollerslev1986}.
These distortions arise because the information-arrival process and the
sampling scheme are mismatched: the clock advances at a constant rate
while market activity does not \cite{AndersenBollerslev1997,Ane_Geman2000}.
In cryptocurrency markets this mismatch is particularly severe, as
Bariviera \emph{et al.}~\cite{Bariviera2017} document significant
long-range dependence in Bitcoin returns that is inconsistent with the
i.i.d.\ assumption implicit in calendar-bar analysis, and
Katsiampa~\cite{Katsiampa2017} shows that Bitcoin volatility exhibits
both short-run and long-run conditional variance components that simple
GARCH models underfit when applied to time-sampled data.
Economically, calendar bars conflate periods of genuine price discovery
with noise-driven drift, obscuring the microstructure signals that
matter for short-horizon prediction and execution
\cite{Kyle1985,GlostenMilgrom1985,Hasbrouck1991}.
Table~\ref{tab:abbrev} lists the abbreviations used throughout this paper.

\begin{table}[!t]
\caption{List of Abbreviations}
\label{tab:abbrev}
\centering
\renewcommand{\arraystretch}{1.2}
\setlength{\tabcolsep}{6pt}
\begin{tabular}{@{}ll@{}}
\toprule
\textbf{Abbreviation} & \textbf{Definition} \\
\midrule
ACF    & Autocorrelation Function \\
ADF    & Augmented Dickey-Fuller test \\
ARCH   & Autoregressive Conditional Heteroskedasticity \\
BDS    & Brock-Dechert-Scheinkman test \\
BTC    & Bitcoin \\
BTCUSDT & Bitcoin / Tether trading pair \\
CV     & Coefficient of Variation \\
EMA    & Exponentially Weighted Moving Average \\
GARCH  & Generalised ARCH \\
i.i.d. & Independent and Identically Distributed \\
JB     & Jarque-Bera statistic \\
KPSS   & Kwiatkowski-Phillips-Schmidt-Shin test \\
LB     & Ljung-Box statistic \\
MDH    & Mixture-of-Distributions Hypothesis \\
ML     & Machine Learning \\
OHLCV  & Open, High, Low, Close, Volume \\
TO     & Timeout (bar closed by duration guard) \\
USDT   & Tether (stablecoin) \\
VPIN   & Volume-Synchronised Probability of Informed Trading \\
VR     & Variance Ratio \\
VWAP   & Volume-Weighted Average Price \\
\bottomrule
\end{tabular}
\end{table}

The theoretical foundation for a superior approach was laid by
Clark~\cite{Clark1973}, who demonstrated that subordinating calendar
time to a directing process tied to market activity produces return
distributions substantially closer to Gaussian.
This insight, grounded in the mixture-of-distributions hypothesis
\cite{Tauchen_Pitts1983,Andersen1996}, motivates the construction of
\emph{information-driven bars}: sampling rules that close a new bar
when the accumulated measure of market activity reaches a target
threshold, rather than when a clock interval expires.
Volume bars close on cumulative traded quantity.
Dollar bars threshold on cumulative dollar turnover (price times
quantity), addressing the non-stationarity of raw volume across an
asset's price history; this is especially important for Bitcoin, whose
price has appreciated by several orders of magnitude since inception,
making raw volume counts non-comparable across time
\cite{LopezdePrado2018}.
Volatility bars close when accumulated realised return movement
reaches a target, directly aligning sampling frequency with the rate
of price discovery in the underlying market.
Range bars and Renko bars impose a minimum price excursion before a
bar is formed, filtering microstructure noise \cite{Kaufman2013}.
Hybrid bars, which condition bar closure on satisfaction of
either a dollar-volume or a volatility criterion (whichever fires first),
represent a further generalisation that captures information arrival
across several dimensions simultaneously.

The systematic study of these alternatives has been substantially
advanced by L\'{o}pez de Prado~\cite{LopezdePrado2018}, whose unified
framework for information-driven bar construction motivates their
application to machine learning pipelines.
An\'{e} and Geman~\cite{Ane_Geman2000} demonstrated theoretically
that subordinating returns to cumulative trade count yields
approximately Gaussian innovations, removing the fat tails that plague
calendar-sampled cryptocurrency return series
\cite{Bariviera2017,Katsiampa2017}.
Easley \emph{et al.}~\cite{Easley_Lopez_OHara2012} connected
dollar-bar frequency to the probability of informed trading, providing
a microstructure grounding for dollar-volume thresholds that is
directly applicable to the continuous, order-flow-driven dynamics of
cryptocurrency exchanges.
The same authors showed empirically that synchronising sampling to
cumulative volume produces series with more uniform bar sizes and
reduced serial dependence compared to fixed calendar intervals
\cite{Easley2012TheVC}; no equivalent controlled comparison
yet exists in the cryptocurrency literature, making this the strongest
available methodological evidence for the approach.

\subsection{The Open Problem: Data Resolution and Bar Construction}

Despite this theoretical and empirical motivation, a critical question
has received insufficient systematic attention:
\emph{does the resolution of the underlying input data, specifically
raw tick-level trades versus pre-aggregated minute bars, materially
alter the properties of the resulting information bars?}

This question is not merely academic.
In practice, researchers frequently construct information bars from the
lowest-resolution data conveniently available (minute OHLCV bars
sourced from exchange APIs) rather than from raw tick data, which is
voluminous and computationally demanding.
When minute OHLCV data is used, per-bar accumulation is necessarily
coarser: dollar volume is approximated as close price times total
minute volume; volatility is captured only at the close-to-close
boundary; and the intra-minute price path (including all reversals,
microstructure events, and partial fills) is irretrievably lost.
The tick-level pipeline, by contrast, processes every individual trade:
dollar volume is an exact summation of price-times-quantity per trade,
realized volatility accumulates as the sum of absolute log-returns
across every consecutive trade pair within the bar, and the bar's high
and low reflect the actual extremes of executed prices.

These are not merely computational differences; they represent
fundamentally distinct operationalisations of the same conceptual bar
type.
A price sequence containing three intra-minute reversals of 0.05\%
that cancel out yields a close-to-close return of approximately zero
and contributes nothing to the minute-based volatility accumulator,
yet contributes 0.15\% to the tick-based realized volatility
accumulator.
Similarly, the minute-based range bar sums per-minute
$(H_i - L_i)/C_i$ spans across successive one-minute buckets; the
tick-based range bar tracks the exact running maximum and minimum of
individual trade prices, closing when their relative distance first
exceeds the target.
For Renko bars, the difference is detection latency: the minute
pipeline can only check the displacement condition at minute
boundaries, potentially overshooting the target brick size; the tick
pipeline detects the crossing event at the precise trade where the
threshold is first breached.

Whether these differences translate into statistically and economically
meaningful divergences in bar-series properties (return distributions,
serial correlation structure, volatility clustering, bar-size
uniformity, and machine learning information content) is an empirical
question that, to the authors' knowledge, has not been addressed with
systematic rigour in the academic literature.

\subsection{Contribution of This Paper}

This paper fills that gap.
Its specific contributions are as follows.

\begin{enumerate}[leftmargin=1.4em,itemsep=3pt,topsep=3pt]
\item \textbf{A frequency-controlled tick-versus-minute comparison.}
  We isolate input data resolution as the sole experimental variable by
  driving two otherwise-identical pipelines from raw Binance aggTrade tick
  data and from pre-aggregated one-minute OHLCV bars, with both calibrated
  through a common adaptive framework and benchmarked against fixed-interval
  calendar bars.
\item \textbf{Six information bar types under one protocol.}
  Dollar, volume, volatility, range, Renko, and hybrid bars are implemented
  and evaluated side by side, rather than the single bar type (typically
  dollar or volume) studied in most prior work.
\item \textbf{Adaptive EMA threshold calibration.}
  Thresholds adapt online through an exponentially weighted moving average
  whose rate is itself governed by a self-monitoring stability routine,
  replacing the static thresholds common in the literature.
\item \textbf{Strictly tick-native construction logic.}
  Each tick-level bar type accumulates its activity signal from individual
  trade prices and sizes (exact within-bar excursions, displacements, and
  dollar volume), ensuring genuine methodological differentiation rather than
  mere resolution upscaling of the minute signal.
\item \textbf{A matched-frequency robustness analysis.}
  We coarsen the tick series to the minute pipeline's bar count to separate
  genuine resolution effects from sampling-frequency artefacts, a confound
  that otherwise contaminates raw cross-resolution comparisons.
\item \textbf{A six-year public cryptocurrency dataset.}
  All results are computed on continuous BTCUSDT USDT-margined perpetual
  futures data spanning January 2020 to December 2025, covering multiple
  volatility regimes, with the full implementation released as replication
  materials.
\end{enumerate}

Both pipelines are evaluated across eight statistical quality
criteria (detailed in Section~\ref{subsec:evaluation}) drawn from the
information-bar literature~\cite{LopezdePrado2018} and the
cryptocurrency microstructure literature.
These criteria assess three theoretical properties: distributional
normality of bar returns, serial independence of the return sequence,
and bar construction quality.
All-return-series stationarity is verified as a prerequisite.

The eventual objective is the application of machine learning methods
to the resulting bar series.
Information-driven bars have been proposed as superior training
substrates for supervised learning models because their more uniform
information content reduces the look-ahead and serial correlation
biases that inflate in-sample performance on calendar-based series
\cite{LopezdePrado2018,Dixon_Halperin_Bilokon2020,
Sezer_Gudelek_Ozbayoglu2020}.

\subsection{Methodological Considerations}
\label{subsec:flags}

Several implementation choices deviate from or extend canonical
formulations and should be noted before proceeding.

\textbf{Adaptive thresholds via EMA.}
Both pipelines adapt bar thresholds online via exponentially weighted
moving averages~\cite{Wilder1978}.
The adaptation rate $\alpha$ is itself calibrated from historical
coefficient-of-variation and regime stability metrics, introducing a
path-dependent feedback loop absent from the static threshold
implementations typically reported in the literature.
Results should be interpreted as properties of the adaptive pipeline
rather than of a fixed-threshold rule.
A consequence is that the observed statistical improvements reflect the
joint effect of activity-based sampling \emph{and} EMA-induced threshold
smoothing; a controlled ablation comparing fixed and adaptive thresholds
would be required to isolate each component's contribution, and
constitutes an important direction for follow-up work.

\textbf{Duration constraints.}
Both pipelines enforce minimum and maximum duration bounds
independently of the activity threshold.
A minimum duration prevents pathologically short bars during
high-frequency bursts; a maximum duration forces closure in
low-activity periods even when the threshold is unmet.
This introduces a hybrid calendar/activity character that must be
accounted for in comparative evaluation.

\textbf{Renko displacement semantics.}
In the minute pipeline, the Renko accumulator is the
\emph{instantaneous} displacement
$|p_t - p_{\mathrm{ref}}| / p_{\mathrm{ref}}$
rather than a running sum, so it can \emph{decrease} on price
reversal.
Classical Renko construction holds bricks once formed and requires a
full additional brick to reverse direction; readers comparing results
to classical Renko studies should account for this distinction
(see Section~\ref{subsec:renko}).

\textbf{Range bar signal divergence.}
The minute pipeline accumulates the sum of per-minute relative ranges
$\sum_i (H_i - L_i)/C_i$; the tick pipeline accumulates the exact
within-bar excursion
$(H_{\mathrm{bar}} - L_{\mathrm{bar}})/p_{\mathrm{open}}$
from individual trade prices.
The minute approximation systematically underestimates true high-low
span in mean-reverting intra-minute environments.

\textbf{Hybrid bar closure logic.}
The hybrid bar closes when \emph{either} the dollar-volume threshold
\emph{or} the volatility threshold is first reached (OR logic), plus
the minimum duration constraint is satisfied.
OR logic is used because AND logic (requiring both thresholds
simultaneously) produces timeout rates exceeding 60\% at the
dollar-volume/volatility correlation ($\rho \approx 0.5$--$0.7$)
characteristic of Bitcoin, effectively converting hybrid bars into
calendar-interval time bars.
OR logic yields approximately 65\% organic close rate at typical
correlation levels, preserving the information-sampling advantage of
non-uniform bars (see Section~\ref{subsec:hybrid}).

\subsection{Paper Organisation}

Section~\ref{sec:literature} reviews the market microstructure and
information-bar literatures.
Section~\ref{sec:methodology} describes the data sources, calibration
framework, and statistical evaluation protocol.
Section~\ref{sec:barconstruction} details the construction logic for
each of the six bar types across both pipelines.
Section~\ref{sec:results} presents the comparative empirical results.
Section~\ref{sec:mlexperiment} reports a downstream machine learning
experiment that tests whether superior statistical bar properties
translate to superior out-of-sample directional prediction performance.
Section~\ref{sec:discussion} interprets the findings and discusses
implications for machine learning applications.
Section~\ref{sec:conclusion} concludes.

\section{Literature Review}
\label{sec:literature}

\subsection{From Calendar Bars to Activity-Based Sampling}
\label{subsec:lit_bars}

The inadequacy of calendar-based sampling as a representation of
financial price processes was identified long before the modern
information-bar literature.
The intellectual lineage of activity-based sampling stretches back to
Bachelier~\cite{Bachelier1900}, who first modelled security prices as a
continuous random walk, the mathematical foundation on which all
subsequent subordination arguments rest.
Working~\cite{Working1934} first demonstrated empirically that
regular-interval sampling of commodity price series induces spurious
serial correlation not present in the underlying price-generating
process.
Osborne~\cite{Osborne1959} showed that stock price changes measured
in transaction time, rather than in calendar time, are substantially
closer to Gaussian, anticipating by more than a decade the
subordination argument that Clark~\cite{Clark1973} would formalise
rigorously.
Samuelson~\cite{Samuelson1965} provided the formal proof that properly
anticipated prices fluctuate randomly, establishing the random walk as
the null hypothesis against which any departure motivated by
activity-based sampling must be measured.
Mandelbrot and Taylor~\cite{MandelbrotTaylor1967} provided an early
empirical demonstration that sampling in transaction time, rather than
calendar time, yields return distributions that better conform to
stable Paretian models, directly prefiguring the subordination argument.
Clark~\cite{Clark1973} subsequently formalised this intuition by modelling
speculative prices as a Brownian motion directed by a stochastic activity
clock tied to cumulative trading volume, proving that subordination removes
the fat tails and serial dependence that afflict calendar-sampled returns.
The mixture-of-distributions hypothesis~\cite{Tauchen_Pitts1983,
Andersen1996} formalised the same insight: the non-normality of
calendar-sampled returns reflects the mixing of observations drawn
from periods with fundamentally different information intensities.

Fama's efficient market hypothesis~\cite{Fama1970} provides the
theoretical benchmark: under full informational efficiency, no
sampling scheme should systematically improve return distributional
properties.
The empirical evidence cited in this section, serial correlation,
fat tails, and volatility clustering in calendar-sampled returns, 
constitutes direct evidence against this benchmark and motivates
the information-bar approach.

These theoretical results motivate a direct operational question:
how should the sampling clock be defined?
An\'{e} and Geman~\cite{Ane_Geman2000} provided the most direct
empirical answer, showing that subordinating equity and foreign
exchange returns to cumulative transaction count yields approximately
Gaussian innovations.
Andersen and Bollerslev~\cite{AndersenBollerslev1997} further
documented that intraday periodicity and volatility persistence
characteristic of calendar-sampled equity returns are substantially
attenuated under activity-based aggregation.
Dacorogna \emph{et al.}~\cite{Dacorogna_etal2001} synthesised these
insights into a unified high-frequency finance framework centred on
\emph{theta-time}, a physical-activity clock that contracts during
periods of low market activity and expands during high activity, 
establishing the methodological vocabulary that motivates the present
paper's adaptive bar-construction approach.

L\'{o}pez de Prado~\cite{LopezdePrado2018} synthesised this body of
work into a unified framework for \emph{information-driven bars},
operationalising four activity measures as sampling clocks: tick
count, cumulative volume, cumulative dollar turnover, and realised
price movement.
Dollar bars are particularly important for assets with large
historical price appreciation, where raw volume counts are
non-stationary: sampling on dollar turnover normalises for both price
level and traded quantity simultaneously.
Easley \emph{et al.}~\cite{Easley_Lopez_OHara2012}
connected dollar-bar frequency to the Volume-Synchronised Probability
of Informed Trading (VPIN), establishing that dollar bars close more
frequently when informed order flow is elevated, thereby embedding a
market-toxicity signal directly in the bar structure.
The same group~\cite{Easley2012TheVC} demonstrated empirically
that volume-clock sampling produces more uniform bar sizes and reduced
serial dependence compared to fixed calendar intervals, providing the
strongest available evidence for the activity-bar framework.

Price-path-based bar types complement the activity-based family.
Range bars close when the within-bar high-low span reaches a
threshold, suppressing microstructure noise and reducing intraday
seasonality~\cite{Kaufman2013}.
Renko bars close on fixed price displacements, yielding series in
which every bar carries equal price information.
Hybrid bars conjoin multiple criteria (typically volume and
realised volatility) and represent a natural but empirically
understudied generalisation.
Despite more than a decade since L\'{o}pez de Prado's formalisation,
the academic literature, to the authors' knowledge, still lacks a
controlled, systematic comparison of these bar types constructed from
tick-level data versus the minute-level OHLCV data that practitioners
routinely use as a lower-cost substitute.
This is the primary gap the present paper addresses.

\subsection{Cryptocurrency Market Structure and Statistical Properties}
\label{subsec:lit_crypto}

Cryptocurrency markets exhibit structural features that make the
calendar-sampling problem especially severe and that simultaneously
make them an unusually clean test bed for activity-based bar
construction.

Bitcoin~\cite{Nakamoto2008} and major cryptocurrencies trade continuously without session
boundaries, removing the market-open and market-close effects that
confound intraday periodicity studies in equity markets.
The statistical properties of Bitcoin returns are extensively
documented and differ markedly from those of traditional assets.
The canonical stylised facts of financial returns, fat tails,
volatility clustering, and near-zero autocorrelation of raw returns
but positive autocorrelation of absolute and squared returns, 
were established for equity markets by Cont~\cite{Cont2001} and
are replicated with greater severity in cryptocurrency markets.
Bariviera \emph{et al.}~\cite{Bariviera2017} establish significant
long-range dependence in Bitcoin returns using the Hurst exponent on
transaction-level data from 2011 to 2017, finding multifractal
structure inconsistent with the i.i.d.\ assumption implicit in
calendar-bar analysis.
Katsiampa~\cite{Katsiampa2017} shows that Bitcoin volatility requires
a component GARCH specification with distinct short-run and long-run
conditional variance dynamics, motivating realised-volatility bars
that adapt to both components in real time.
Urquhart~\cite{Urquhart2016} documents statistically significant
return autocorrelation in Bitcoin using variance ratio and BDS tests,
directly confirming the serial dependence that information bars are
theoretically designed to attenuate; Nadarajah and
Chu~\cite{Nadarajah_Chu2017} corroborate this finding using eight
alternative test statistics on the same data, confirming that the
result is not test-dependent.
Baur \emph{et al.}~\cite{Baur_etal2018} establish that Bitcoin
functions primarily as a speculative asset with return properties
largely orthogonal to traditional asset classes.
Brauneis and Mestel~\cite{Brauneis_Mestel2018} test Bitcoin price
discovery efficiency and find that sampling frequency materially
affects efficiency metrics, providing direct support for the premise
that the choice of bar construction method is consequential for
cryptocurrency research.
Tran and Leirvik~\cite{Tran_Leirvik2020} show that Bitcoin's
market efficiency is time-varying rather than fixed, implying that
any bar construction framework must adapt its parameters to the
prevailing market regime, precisely the motivation for the adaptive
EMA calibration framework adopted in Section~\ref{subsec:calibration}.

At the market structure level, Makarov and
Schoar~\cite{Makarov_Schoar2020} document large and persistent
arbitrage deviations across cryptocurrency exchanges using tick-level
transaction data, highlighting the central role of order flow in
cryptocurrency price formation.
Griffin and Shams~\cite{Griffin_Shams2020} demonstrate that tether
issuance was used to manipulate Bitcoin prices, further establishing
that cryptocurrency order flow operates through mechanisms distinct
from those of traditional asset markets.
Easley \emph{et al.}~\cite{Easley_OHara_2024} apply VPIN,
Roll spread estimates~\cite{Roll1984}, and the Amihud illiquidity
ratio to Binance data for five major cryptocurrencies, finding that
microstructure measures have significant predictive power for
short-horizon price dynamics.
Liu \emph{et al.}~\cite{Liu_Tsyvinski_Wu2022} and
Liu and Tsyvinski~\cite{Liu_Tsyvinski2021} establish that
cryptocurrency return cross-sections are driven by asset-class-specific
market, size, and momentum factors with no meaningful exposure to
traditional equity or macroeconomic risk factors, implying that the
information environment of cryptocurrency markets operates through
mechanisms that standard calendar-bar analysis is poorly suited to
capture.
Biais \emph{et al.}~\cite{Biais_etal2023} provide a structural
equilibrium model of Bitcoin pricing that grounds these asset-class
dynamics in the economics of proof-of-work mining, establishing a
theoretical foundation for the price-formation process that the
present paper's information bars are designed to sample.

\subsection{Information Bars, Machine Learning, and Open Problems}
\label{subsec:lit_ml}

The intersection of information-driven bar construction and machine
learning has attracted growing empirical attention.
L\'{o}pez de Prado~\cite{LopezdePrado2018} argues that more uniform
information content per bar reduces the serial correlation and
look-ahead biases that inflate in-sample performance of models
trained on calendar-sampled series, a claim that applies with
particular force to cryptocurrency markets given the documented
autocorrelation of Bitcoin returns~\cite{Urquhart2016}.
Sezer \emph{et al.}~\cite{Sezer_Gudelek_Ozbayoglu2020}
survey over two hundred deep learning studies in financial forecasting
and identify input data representation as one of the most consequential
and least systematically studied design choices.
Lim and Zohren~\cite{Lim_Zohren2021} extend this analysis to
transformer-based architectures and emphasise that the stationarity
of the time-step information content, directly measured by bar-size
CV in the present paper's evaluation framework, is a prerequisite
for consistent learning across long sequences.
Fang \emph{et al.}~\cite{Fang_etal2022} survey 146 cryptocurrency
trading papers and note that the overwhelming majority use
calendar-sampled minute or daily data; alternative bar types are
rarely considered.

The most directly related recent work is
Gr{\k{a}}dzki \emph{et al.}~\cite{Gradzki_etal2025}, who apply
volume bars, dollar bars, range bars, and the CUSUM filter to Binance
tick data for Bitcoin and Ethereum (2018--2023), combining them with
the Triple Barrier labelling method and deep learning classifiers.
Their results confirm that activity-based bars produce statistically
distinct training environments from time bars and that bar choice
materially affects ML strategy performance.
However, their study does not compare tick-level bar construction
against minute-level construction of the same bar types, does not
include volatility bars or hybrid bars, and does not conduct a
systematic statistical evaluation against the de~Prado quality
criteria.
These are precisely the contributions of the present paper.

Two further gaps remain open.
First, no study has evaluated the timeout percentage (the fraction of bars
closed by the maximum duration guard rather than by the activity signal)
as an explicit quality metric, despite it being
a direct diagnostic of calibration failure.
Second, the nonlinear dependence structure of information-bar
returns, as measured by the BDS test~\cite{BDS1996}, which detects
stochastic dependence structures invisible to standard autocorrelation
tests by examining correlation-dimension statistics across multiple
embedding dimensions, has not been studied in the cryptocurrency
context.
Standard autocorrelation tests may understate serial dependence when
returns exhibit the nonlinear dynamics documented for
Bitcoin~\cite{Katsiampa2017}.
The evaluation framework adopted here addresses both gaps.

\section{Data and Experimental Setup}
\label{sec:methodology}

\subsection{Data Sources}
\label{subsec:data}

Both pipelines are applied to data from the Binance BTCUSDT
USDT-margined perpetual futures market~\cite{Binance2024}, the
continuously traded Bitcoin/Tether perpetual swap contract on Binance,
the largest cryptocurrency derivatives exchange by volume.
USDT-margined perpetual futures are the primary vehicle for leveraged
Bitcoin price discovery on centralised exchanges, exhibit the highest
liquidity and order-flow activity of any Bitcoin-denominated instrument,
and track spot prices closely via continuous funding-rate arbitrage.
The statistical properties of perpetual futures returns (fat tails,
volatility clustering, and serial dependence in absolute returns) mirror
those documented for spot Bitcoin in the literature reviewed in
Section~\ref{subsec:lit_crypto}, making perpetual futures a natural
and representative test bed for information bar construction~\cite{Easley_OHara_2024}.
The study period spans 1 January 2020 to 31 December 2025, encompassing
2,192 calendar days of continuous 24-hour trading.
Two parallel datasets covering the identical period are used,
differing only in their temporal resolution.

\textbf{Tick-level data.}
The primary dataset consists of Binance aggTrade records for the
BTCUSDT USDT-margined perpetual futures contract.
Each aggTrade record represents a single market-aggressor event
aggregated across fills that share the same price, direction, and
millisecond timestamp, and contains: trade price ($p$), aggregated
quantity ($q$), Unix timestamp in milliseconds ($t$), and a buyer-maker
flag ($b$) indicating whether the buyer or seller was the aggressor.
Dollar volume per record is computed exactly as $p \times q$.
The dataset is processed sequentially to handle its volume, with
all bar accumulators maintained continuously across the full period
to ensure no activity signal is lost between processing batches.

\textbf{Minute OHLCV data.}
The secondary dataset consists of Binance one-minute OHLCV bars
for the same BTCUSDT perpetual futures contract and period.
Each record contains open, high, low, close prices and total volume
for the corresponding one-minute window.
Dollar volume per minute is approximated as $\text{close} \times \text{volume}$,
which is the standard approximation available from pre-aggregated data.
This approximation discards the intra-minute price path and introduces
a systematic bias relative to the exact tick-level computation.

\textbf{Calibration period.}
Both pipelines share a common calibration procedure using the most
recent $L = 14$ calendar days of data prior to the construction period.
This window is chosen to span at least two full weekly cycles of
Bitcoin trading activity while remaining within a single volatility
regime; shorter windows risk under-representing the distribution of
market activity, while longer windows may incorporate multiple
structural breaks.
The calibration framework (lookback window length, EMA adaptation rate,
and duration bounds) is derived from minute OHLCV data for both pipelines
(see Section~\ref{subsec:calibration} below).
For the tick pipeline, the initial activity threshold is subsequently
replaced with a tick-native equivalent computed from the same lookback
window of tick data: exact $\sum p_i q_i$ for dollar bars, realised
volatility $\sum\bigl|\log(p_i/p_{i-1})\bigr|$ for volatility bars, and exact
running high-low span for range bars.
This design ensures that the calibration framework is shared while the
activity signal resolution differs, isolating signal resolution as the
sole experimental variable.

\subsection{Calibration Framework}
\label{subsec:calibration}

A core design principle of both pipelines is that bar construction
parameters adapt to prevailing market conditions rather than relying
on fixed thresholds.
Static thresholds appropriate for one market regime (low volatility,
high liquidity) become systematically miscalibrated under a different
regime.
The calibration framework addresses this through a two-stage procedure.

\textbf{Stage 1: Historical initialisation.}
For each bar type, an initial market analysis procedure
processes the $L$-day calibration window of minute OHLCV data to
derive three sets of parameters:
\begin{enumerate}
    \item The initial activity threshold (target dollar volume,
    target volume, target volatility, target brick size, or target
    range), computed as the median daily activity measure divided by
    the target bars-per-day.
    \item The EMA adaptation rate $\alpha \in [\alpha_{\min},
    \alpha_{\max}]$, calibrated from the coefficient of variation of
    daily activity and a rolling regime stability measure.
    \item Duration bounds $[d_{\min}, d_{\max}]$ in minutes,
    derived from the estimated bar duration and the frequency
    target.
\end{enumerate}
For the tick pipeline, Stage 1 is followed by a tick-native threshold
replacement: the minute-derived threshold is discarded and replaced
with the median daily tick-native activity measure divided by the same
target bars-per-day.
For volatility bars, this is
\begin{equation*}
  \tilde{\sigma} = \frac{\operatorname{median}\!\left(
      \sum_{i} \left|\log\!\left(\tfrac{p_i}{p_{i-1}}\right)\right|_{\!\text{day}}
    \right)}{\bar{n}},
\end{equation*}
where the sum is taken over all consecutive tick pairs within each
calendar day and $\bar{n}$ is the target bars-per-day.
For range bars, the tick-native threshold is
\begin{equation*}
  \tilde{r} = \frac{\operatorname{median}\!\left(
    \tfrac{H_{\text{day}} - L_{\text{day}}}{p_{\text{first}}}
    \right)}{\bar{n}},
\end{equation*}
where $H_{\text{day}}$ and $L_{\text{day}}$ are the day's exact high
and low from individual trade prices and $p_{\text{first}}$ is the
day's first trade price.
For dollar bars, the tick-native threshold is
\begin{equation*}
  \tilde{d} = \frac{\operatorname{median}\!\left(
    \sum_{i} p_i q_i{}_{\,\text{day}}\right)}{\bar{n}}.
\end{equation*}

\textbf{Stage 2: Online EMA adaptation.}
After each bar closes, the activity threshold is updated via an
exponentially weighted moving average:
\begin{equation}
    \hat{\theta}_{n+1} = (1 - \alpha)\,\hat{\theta}_n +
    \alpha \cdot \min\!\left(s_n,\, 2\hat{\theta}_n\right),
    \label{eq:ema_update}
\end{equation}
where $\hat{\theta}_n$ is the current threshold and $s_n$ is the
actual bar size (accumulated activity).
The cap at $2\hat{\theta}_n$ is an original design choice introduced
to prevent a single extreme-activity event (flash crash, news spike)
from permanently inflating the threshold: without this cap, a
single bar of size $10\hat{\theta}_n$ would shift the EMA-adapted
target far above the prevailing activity level, producing dozens of
subsequent timeout bars.
The adaptation rate $\alpha$ is itself updated periodically via a
self-monitoring routine that computes the coefficient of variation of
the recent threshold history and maps it monotonically to
$[\alpha_{\min},\alpha_{\max}]$: high threshold variability (unstable
market conditions) raises $\alpha$ for faster adaptation, while low
variability (stable conditions) lowers $\alpha$ for smoother tracking.

\textbf{Why an EMA?}
The threshold update in \eqref{eq:ema_update} is deliberately the simplest
estimator that satisfies the three requirements of an online bar-sampling
clock, and we considered several alternatives before adopting it.
A \emph{rolling median} or fixed-window \emph{quantile} is more robust to
outliers, but requires buffering the last $w$ bar sizes and re-sorting them at
every close; at the tick pipeline's tens of millions of updates this is both
memory- and compute-heavier, and it introduces a hard window edge that
discards information abruptly rather than decaying it.
An \emph{exponentially weighted variance} estimator targets the second moment
of activity, whereas the quantity a bar-closing rule must track is the
\emph{level} of typical activity; variance adaptation would additionally have
to be paired with a level estimator, reintroducing the EMA we already use.
A \emph{Kalman filter} is the natural optimal choice under an explicit
linear-Gaussian state-space model, but activity sizes are strongly
non-Gaussian and heavy-tailed (Section~\ref{sec:results}), so the Kalman
optimality guarantee does not hold, and the filter would demand process- and
observation-noise covariances that are themselves unknown and regime-varying.
The EMA, by contrast, is a constant-memory $O(1)$-per-bar recursion that
decays old information smoothly, exposes a single interpretable rate parameter
$\alpha$, and (because $\alpha$ is itself adapted from the recent
coefficient of variation) already recovers the main practical benefit of a
Kalman filter (faster tracking when the series is volatile, smoother tracking
when it is calm) without its distributional assumptions.
The bounded update $\min(s_n, 2\hat{\theta}_n)$ supplies the outlier
robustness that motivates the median alternative.
A controlled ablation quantifying the sensitivity of bar quality to the choice
of adaptation estimator is left to future work.

\textbf{Duration constraints.}
Both pipelines enforce a minimum duration $d_{\min}$ and a maximum
duration $d_{\max}$ per bar, independently of the activity threshold.
A bar that has not yet accumulated sufficient activity is held open
until $d_{\min}$ has elapsed; a bar that has not reached its threshold
by $d_{\max}$ is force-closed regardless.
Force-closed bars are flagged as timeout bars.
The timeout percentage, defined as the fraction of bars closed by
$d_{\max}$ rather than by the activity signal, is reported as a
calibration diagnostic: a timeout rate exceeding 10\% indicates that
the threshold is systematically unachievable under current market
conditions, and the resulting bars carry no genuine microstructure
information~\cite{LopezdePrado2018}.

\subsection{Evaluation Framework}
\label{subsec:evaluation}

Each bar type is evaluated across eight statistical criteria drawn
from the information-bar and financial econometrics literatures.
A fixed-interval time-bar series with the same approximate frequency
as the corresponding information bars serves as the null-hypothesis
baseline: if an information bar cannot outperform a calendar clock on
most criteria, it is not extracting structure the clock cannot already
provide~\cite{LopezdePrado2018}.

Table~\ref{tab:criteria} summarises the eight criteria, their
direction of preference, their null hypotheses, and the primary
references from which they are drawn.
The criteria are organised into three conceptual groups reflecting
the three theoretical claims of information-bar construction:
distributional normality (from the mixture-of-distributions hypothesis),
serial independence (the primary de~Prado criterion), and information
density.

\textbf{Primary metrics versus secondary diagnostics.}
Although eight criteria are computed, they are not weighted equally in
our interpretation, and we designate a small primary set to which the
central claims of the paper are anchored.
The \emph{primary} metrics are the variance-ratio deviation
$|\mathrm{VR}(4){-}1|$, the lag-1 autocorrelation magnitude
$|\mathrm{AC}(1)|$, and excess kurtosis: the first two are effect-size
measures of random-walk conformance (the core property information bars
are designed to deliver), and the third is the standard scalar summary of
tail heaviness under the mixture-of-distributions hypothesis.
All remaining quantities (Jarque-Bera, Ljung-Box $p$-values, bar-size CV,
Shannon entropy, timeout percentage, and the supplementary BDS, ARCH-LM,
KS, and Mann-Whitney diagnostics) are treated as \emph{secondary}: they
characterise, corroborate, or contextualise the primary findings but do
not by themselves establish them.
This ordering matters because, as discussed next, the secondary
$p$-value-based tests are of limited discriminating value at the sample
sizes studied.

\textbf{Effect sizes over $p$-values at large $N$.}
The bar series analysed here contain from roughly thirty thousand to over
two hundred thousand observations.
At such sample sizes, any economically negligible departure from a null
hypothesis is detected with overwhelming statistical significance: the
Ljung-Box and Jarque-Bera tests return $p \approx 0$ for essentially every
series, including those whose autocorrelation is economically trivial.
A rejection therefore conveys little about \emph{how far} a series departs
from a random walk.
For this reason we base comparative claims on effect-size magnitudes (the
value of $|\mathrm{AC}(1)|$ and the deviation $|\mathrm{VR}(4){-}1|$) rather
than on the binary outcome of the associated significance test, and we
report Ljung-Box $p$-values chiefly to document that the sample-size regime
renders them uniformly rejecting rather than to rank bar types.

\textbf{Descriptive, not inferential, use of the tests.}
The evaluation applies statistical tests for eight criteria across six
bar types and three series (minute, tick, time bar), producing 48
primary comparisons.
The tests are used descriptively rather than inferentially: the objective
is characterisation of cross-bar patterns through comparable, standardised
summary statistics, not binary accept/reject decisions at a controlled
family-wise error rate.
Because no formal decision procedure is being run, no multiple-testing
correction (such as Holm-Bonferroni or Benjamini-Hochberg) is applied or
required; correspondingly, individual $p$-values near conventional
thresholds carry no special status.
Results should instead be read in their totality: the consistent
directional patterns documented across multiple bar types and multiple
criteria are more informative than any individual test outcome.

\begin{table}[tp]
\caption{Statistical Evaluation Criteria}
\label{tab:criteria}
\centering
\renewcommand{\arraystretch}{1.3}
\begin{tabular}{@{}p{2.35cm}p{1.35cm}p{2.15cm}l@{}}
\toprule
\textbf{Criterion} & \textbf{Direction} & \textbf{Null hypothesis ($H_0$)} & \textbf{Ref.} \\
\midrule
\multicolumn{4}{@{}l}{\textit{Distributional Normality}} \\[1pt]
Excess kurtosis       & $|k| \to 0$    & Returns Gaussian    & \cite{Mandelbrot1963,Bariviera2017} \\
Jarque-Bera statistic & Lower          & Joint normality     & \cite{JarqueBera1980} \\[4pt]
\multicolumn{4}{@{}l}{\textit{Serial Independence}} \\[1pt]
Ljung-Box $p$(10)     & Higher         & No autocorrelation  & \cite{LjungBox1978} \\
$|\mathrm{VR}(4){-}1|$ & $\to 0$      & Random walk         & \cite{LoMacKinlay1988} \\
$|\mathrm{AC}(1)|$    & $\to 0$      & Serial independence & \cite{LoBio1990,Urquhart2016} \\[4pt]
\multicolumn{4}{@{}l}{\textit{Bar Construction Quality}} \\[1pt]
Bar-size CV           & Lower          & Uniform bar sizes   & \cite{LopezdePrado2018} \\
Shannon entropy       & Higher         & Max.\ information   & \cite{Shannon1948} \\
Timeout percentage    & Lower          & Signal-driven close & \cite{LopezdePrado2018} \\[4pt]
\multicolumn{4}{@{}l}{\textit{Stationarity (prerequisite, not scored)}} \\[1pt]
ADF $p$-value         & ${<}\,0.05$    & Unit root absent    & \cite{DickeyFuller1979} \\
KPSS $p$-value        & ${>}\,0.05$    & Stationarity holds  & \cite{KPSS1992} \\
\bottomrule
\end{tabular}
\end{table}

In addition to the primary scorecard, the following supplementary
diagnostics are reported for each bar type.

\textbf{Autocorrelation profile.}
Lag-1 autocorrelation~\cite{LoBio1990,Urquhart2016} and the Ljung-Box
statistic at 20 lags~\cite{LjungBox1978} supplement the 10-lag primary
criterion.
Both are expected to be near zero for well-constructed information bars
and are known to be elevated in calendar-sampled Bitcoin
returns~\cite{Urquhart2016}.

\textbf{Nonlinear dependence.}
The BDS test at embedding dimensions 2 and 3~\cite{BDS1996} detects
nonlinear serial dependence invisible to autocorrelation tests.
Bitcoin returns exhibit nonlinear GARCH-type dynamics~\cite{Katsiampa2017}
that may persist in information bars even after linear autocorrelation
is accounted for; the BDS test is therefore a necessary complement to
the Ljung-Box criterion.

\textbf{Volatility clustering.}
The ARCH-LM test~\cite{Engle1982} and the Ljung-Box test on squared
returns assess whether volatility clustering persists in the bar
series.
Persistent volatility clustering is expected in any cryptocurrency
series~\cite{Katsiampa2017} and does not constitute a failure of the
information-bar framework; it is reported to characterise the residual
conditional heteroskedasticity in each pipeline.

\textbf{Distribution comparisons.}
The two-sample Kolmogorov-Smirnov test and the Mann-Whitney U test
compare the full return distributions across all three series (minute
vs.\ tick, minute vs.\ time bar, tick vs.\ time bar).
Cohen's $d$ quantifies the effect size of any distributional
differences.
The cross-series comparisons serve a diagnostic function: if an
information bar series is statistically indistinguishable from the
time-bar baseline, it is not extracting structure beyond what the
clock already provides.

\textbf{Tick-level exclusive features.}
Three features are available exclusively from the tick pipeline and
are not computable from minute OHLCV data: VWAP per bar, computed
as $\sum(p_i q_i) / \sum q_i$ over all ticks in the bar; buy-sell
imbalance, defined as $(V_{\text{buy}} - V_{\text{sell}}) /
(V_{\text{buy}} + V_{\text{sell}})$ using dollar volumes; and tick
count per bar.
These features are directly relevant to the informed-order-flow
literature~\cite{Easley_Lopez_OHara2012} and represent additional
information content available to downstream machine learning models
trained on tick-level bars.

\FloatBarrier
\section{Construction of Information Bars}
\label{sec:barconstruction}

Both pipelines share a common architectural pattern.
Each bar type maintains a running accumulator of its activity signal
across successive observations (minutes or ticks), closing a new bar
when the accumulated signal first reaches the adaptive threshold
$\hat{\theta}_n$ defined in Section~\ref{subsec:calibration}, subject
to the duration bounds $[d_{\min}, d_{\max}]$.
At bar closure, all standard OHLCV fields are recorded alongside
the accumulated bar size, bar return, and (for the tick pipeline)
VWAP, tick count, and buy-sell imbalance.
Table~\ref{tab:signals} provides a concise comparison of the
activity signal for each bar type across the two pipelines.
The subsections that follow describe the construction logic and
signal definitions in detail.

\begin{table}[!t]
\caption{Activity Signal Definitions by Bar Type and Pipeline}
\label{tab:signals}
\centering
\renewcommand{\arraystretch}{1.45}
\setlength{\tabcolsep}{5pt}
\begin{tabular}{@{}lll@{}}
\toprule
\textbf{Bar Type} & \textbf{Minute Pipeline} & \textbf{Tick Pipeline} \\
\midrule
Dollar     & $\Delta_m^{\mathrm{dv}} = c_m \cdot v_m$                          & $\Delta_i^{\mathrm{dv}} = p_i \cdot q_i$                          \\[2pt]
Volume     & $\Delta_m^{\mathrm{qty}} = v_m$                                    & $\Delta_i^{\mathrm{qty}} = q_i$                                    \\[2pt]
Volatility & $\Delta_m^{\mathrm{ctc}} = |c_m - c_{m-1}|/c_{m-1}$              & $\Delta_i^{\mathrm{rv}} = |\log(p_i/p_{i-1})|$                    \\[2pt]
Range      & $\Delta_m^{\mathrm{rng}} = (h_m - l_m)/c_m$                       & $\delta_n = (H_b - L_b)/p_{\mathrm{open}}$          \\[2pt]
Renko      & $\delta_m = |c_m - p_{\mathrm{ref}}|/p_{\mathrm{ref}}$ & $\delta_i = |p_i - p_{\mathrm{ref}}|/p_{\mathrm{ref}}$ \\[2pt]
Hybrid     & $\Delta_m^{\mathrm{dv}}$ \textbf{OR} $\Delta_m^{\mathrm{ctc}}$  & $\Delta_i^{\mathrm{dv}}$ \textbf{OR} $\Delta_i^{\mathrm{rv}}$   \\
\bottomrule
\multicolumn{3}{@{}p{\columnwidth}@{}}{\footnotesize
  $c_m$, $h_m$, $l_m$, $v_m$: close, high, low, volume of minute $m$.
  $p_i$, $q_i$: price and quantity of trade $i$.
  $H_b$, $L_b$: running bar high and low (tick pipeline).
  $p_{\mathrm{open}}$, $p_{\mathrm{ref}}$: bar opening price.
  $\Delta$: increment added to running accumulator $S_n$;
  bar closes when $S_n \geq \hat{\theta}$.
  $\delta$: instantaneous displacement (not accumulated);
  bar closes when $\delta \geq \hat{\theta}$.
  dv\,=\,dollar volume; ctc\,=\,close-to-close volatility;
  rv\,=\,realised (tick-native) volatility; qty\,=\,quantity.
  Hybrid: \textbf{OR}: bar closes on whichever signal fires first.}
\end{tabular}
\end{table}

\subsection{Dollar Bars}
\label{subsec:dollar}

A dollar bar closes when the cumulative dollar turnover within the
bar reaches the adaptive threshold.

\textbf{Minute pipeline.}
At each one-minute interval, the dollar-volume increment is
approximated as the product of the closing price and the total
traded volume for that minute:
\begin{equation}
    \Delta_m^{\mathrm{dv}} = c_m \cdot v_m,
    \label{eq:dollar_minute}
\end{equation}
where $c_m$ is the closing price and $v_m$ is the total quantity
traded during minute $m$.
The bar accumulator is the running sum
$S_n = \sum_{m=1}^{n} \Delta_m^{\mathrm{dv}}$, and the bar closes
at the first minute $n^*$ such that $S_{n^*} \geq \hat{\theta}$.

\textbf{Tick pipeline.}
The dollar-volume increment per trade is computed exactly as:
\begin{equation}
    \Delta_i^{\mathrm{dv}} = p_i \cdot q_i,
    \label{eq:dollar_tick}
\end{equation}
where $p_i$ and $q_i$ are the executed price and quantity of the
$i$-th trade.
The accumulator $S_n = \sum_{i=1}^{n} \Delta_i^{\mathrm{dv}}$
advances with every trade rather than at one-minute boundaries.
Additional outputs available only from the tick pipeline include
the volume-weighted average price
$\mathrm{VWAP} = \sum_i(p_i q_i) / \sum_i q_i$
and the buy-sell imbalance
$(V_{\mathrm{buy}} - V_{\mathrm{sell}}) /
(V_{\mathrm{buy}} + V_{\mathrm{sell}})$,
where $V_{\mathrm{buy}}$ and $V_{\mathrm{sell}}$ are the dollar
volumes attributable to buyer-initiated and seller-initiated trades
respectively, as identified by the aggressor-side flag in each
trade record.

\textbf{Signal comparison.}
Equations~\eqref{eq:dollar_minute} and~\eqref{eq:dollar_tick}
reveal the fundamental resolution difference: the minute signal
uses a single closing price as a proxy for all intra-minute
execution prices, whereas the tick signal prices every trade
exactly.
The approximation error is proportional to intra-minute price
volatility and is largest during fast-moving markets where the
closing price deviates substantially from the average transaction
price.

\subsection{Volume Bars}
\label{subsec:volume}

A volume bar closes when the cumulative traded quantity reaches the
adaptive threshold, isolating supply and demand dynamics from price
level.

\textbf{Minute pipeline.}
The per-minute quantity increment is:
\begin{equation}
    \Delta_m^{\mathrm{qty}} = v_m,
    \label{eq:volume_minute}
\end{equation}
accumulated as $S_n = \sum_{m=1}^{n} \Delta_m^{\mathrm{qty}}$,
where $v_m$ is the total traded quantity during minute $m$.

\textbf{Tick pipeline.}
The per-trade quantity increment is:
\begin{equation}
    \Delta_i^{\mathrm{qty}} = q_i,
    \label{eq:volume_tick}
\end{equation}
accumulated as $S_n = \sum_{i=1}^{n} \Delta_i^{\mathrm{qty}}$.
Since total minute volume equals the sum of within-minute trade
quantities, Equations~\eqref{eq:volume_minute}
and~\eqref{eq:volume_tick} are arithmetically equivalent: both
accumulate the exact cumulative traded quantity over the bar's
duration.
The practical difference is temporal resolution: the tick pipeline
closes a bar at the precise trade where the threshold is first
crossed, whereas the minute pipeline can only check the closure
condition at one-minute boundaries, introducing a detection lag of
up to 59 seconds.

\subsection{Volatility Bars}
\label{subsec:volatility}

A volatility bar closes when the accumulated price movement reaches
the adaptive threshold, aligning the sampling clock directly with
the rate of price discovery.
This bar type exhibits the largest signal difference between the two
pipelines.

\textbf{Minute pipeline.}
The per-minute close-to-close volatility increment is:
\begin{equation}
    \Delta_m^{\mathrm{ctc}} =
    \frac{\left|c_m - c_{m-1}\right|}{c_{m-1}},
    \label{eq:vol_minute}
\end{equation}
where $c_m$ and $c_{m-1}$ are the closing prices of consecutive
minutes.
The bar accumulator $S_n = \sum_{m=1}^{n} \Delta_m^{\mathrm{ctc}}$
is the total absolute close-to-close price movement observed across
minutes within the bar.

\textbf{Tick pipeline.}
The tick-native signal is the realised volatility accumulated from
every consecutive trade pair within the bar:
\begin{equation}
    \Delta_i^{\mathrm{rv}} =
    \left|\log\!\left(\frac{p_i}{p_{i-1}}\right)\right|,
    \label{eq:vol_tick}
\end{equation}
and the bar accumulator is
$S_n = \sum_{i=1}^{n} \Delta_i^{\mathrm{rv}}$~\cite{Andersen_etal2001,Andersen_etal2003}.
This signal captures every intra-minute price reversal.
A one-minute interval containing three round-trip oscillations of
$0.05\%$ each contributes $3 \times 0.05\% = 0.15\%$ to the tick
accumulator but only the net close-to-close return (approximately
zero) to the minute accumulator.
The threshold for the tick pipeline is calibrated from the median
daily realised volatility over the 14-day lookback window, replacing
the minute-derived close-to-close target with a tick-native
equivalent of the same conceptual quantity.

\textbf{Signal comparison.}
Equation~\eqref{eq:vol_minute} measures price movement as a sequence
of one-per-minute steps; Equation~\eqref{eq:vol_tick} measures the
exact length of the continuous price path at trade resolution.
The close-to-close approximation is a strict lower bound on the true
intra-bar path length: intra-minute reversals that cancel out
contribute zero to~\eqref{eq:vol_minute} but positive increments
to~\eqref{eq:vol_tick}.
The gap between the two signals is largest in mean-reverting
intra-minute environments and smallest in sustained trending
conditions~\cite{Andersen_etal2003}.

\subsection{Range Bars}
\label{subsec:range}

A range bar closes when the price excursion within the bar reaches
the adaptive threshold.
The two pipelines operationalise this concept in structurally
different ways.

\textbf{Minute pipeline.}
The per-minute range increment is the relative high-low span of
each one-minute interval:
\begin{equation}
    \Delta_m^{\mathrm{rng}} =
    \frac{h_m - l_m}{c_m},
    \label{eq:range_minute}
\end{equation}
where $h_m$, $l_m$, and $c_m$ are the high, low, and closing price
of minute $m$.
The bar accumulator $S_n = \sum_{m=1}^{n} \Delta_m^{\mathrm{rng}}$
is a running sum of per-minute relative spans and is therefore
sensitive to the ordering of intra-minute price moves only through
their impact on minute-level high and low prices.

\textbf{Tick pipeline.}
The tick-native signal is the instantaneous relative price excursion
of the bar computed from individual trade prices:
\begin{equation}
    \delta_n =
    \frac{H_b^{(n)} - L_b^{(n)}}{p_{\mathrm{open}}},
    \label{eq:range_tick}
\end{equation}
where $H_b^{(n)}$ and $L_b^{(n)}$ are the running maximum and
minimum of all trade prices within the current bar, and
$p_{\mathrm{open}}$ is the opening trade price of the bar.
The bar closes when $\delta_n$ first reaches the adaptive threshold.
Unlike the minute accumulator, which sums per-minute spans, this
signal measures the width of the actual price excursion from exact
trade prices and advances only when a new extreme (new bar high or
new bar low) is established.
A price reversal that does not set a new extreme leaves $\delta_n$
unchanged.

\textbf{Signal comparison.}
Equations~\eqref{eq:range_minute} and~\eqref{eq:range_tick} measure
structurally different quantities.
The minute signal accumulates a running sum of per-minute high-low
spans; the tick signal tracks the width of the overall price band
from bar open to the current extreme.
These two quantities are not monotonically related: the minute sum
can exceed the tick excursion when multiple overlapping intra-minute
ranges are summed, and can fall below it when price drifts steadily
in one direction across minutes without establishing large
intra-minute swings.
The two signals therefore select bars on different structural
criteria and are not directly comparable as approximations of the
same underlying quantity.

\subsection{Renko Bars}
\label{subsec:renko}

A Renko bar closes when the price has moved a minimum relative
distance from the bar's reference price, filtering short-term
oscillations and isolating sustained directional moves.

\textbf{Minute pipeline.}
At each one-minute interval, the bar's instantaneous displacement
from its reference price is computed as:
\begin{equation}
    \delta_m =
    \frac{\left|c_m - p_{\mathrm{ref}}\right|}{p_{\mathrm{ref}}},
    \label{eq:renko_minute}
\end{equation}
where $p_{\mathrm{ref}}$ is the opening price of the current bar
and $c_m$ is the current minute's closing price.
The bar closes when $\delta_m \geq \hat{\theta}$ and the minimum
duration constraint is satisfied.

A defining characteristic of this implementation is that
$\delta_m$ is an instantaneous displacement rather than a
running sum.
If price moves away from $p_{\mathrm{ref}}$ and then partially
reverses, $\delta_m$ decreases.
This diverges from the classical Renko convention, in which a brick
is registered once and cannot be unwound; in the classical
formulation, a reversal of the full brick size is required to open
a brick in the opposite direction.
The present implementation therefore closes bars less frequently
than classical Renko during oscillatory markets, producing higher
timeout rates in low-trend regimes, and should be interpreted
accordingly.
To avoid confusion with the classical definition, some practitioners
may prefer to describe this bar type as \emph{displacement bars}:
bars that close when price has sustained a minimum relative excursion
from its bar-open reference, with EMA-adaptive sizing and duration
guards, but without the brick-registration and reversal-threshold
mechanics of classical Renko construction.
This paper retains the ``Renko'' label for continuity with the
broader information-bar literature~\cite{LopezdePrado2018,Kaufman2013}
while explicitly noting the definitional departure in
Section~\ref{subsec:flags}.

\textbf{Tick pipeline.}
The tick-level displacement signal uses the same formula applied
to individual trade prices rather than minute closes:
\begin{equation}
    \delta_i =
    \frac{\left|p_i - p_{\mathrm{ref}}\right|}{p_{\mathrm{ref}}},
    \label{eq:renko_tick}
\end{equation}
where $p_{\mathrm{ref}}$ is the opening price of the current bar
and $p_i$ is the price of the $i$-th trade.
The bar closes at the first trade where $\delta_i \geq \hat{\theta}$
and the minimum duration is satisfied.
Because the threshold is checked at every trade, the tick pipeline
detects brick-sized moves at the precise trade where the threshold
is first breached rather than at the next minute boundary.
This detection resolution, combined with the higher frequency of
threshold crossings inherent in tick-level data, produces
substantially more bars than the minute pipeline over the same
period, reflecting microstructure-level price displacements rather
than the sustained directional moves that the minute pipeline
requires.
The empirical bar counts for the 2020--2025 study period are reported in
Section~\ref{sec:results}.

\textbf{Signal comparison.}
Equations~\eqref{eq:renko_minute} and~\eqref{eq:renko_tick} share
the same displacement formula but differ in the observation on which
it is evaluated: minute-level closing prices versus individual trade
prices.
The consequence is not merely resolution: the tick pipeline can
detect a threshold crossing within the first second of a sustained
move, whereas the minute pipeline must wait until the next
minute close, potentially by which time price has already reversed.

\textbf{Methodological note.}
The large difference in bar counts between the two Renko pipelines
indicates that the tick and minute series are not directly comparable
in the same sense as dollar, volume, or volatility bars.
The two pipelines detect Renko events at fundamentally different
resolution scales, and results are interpreted with this distinction
in mind.
Empirical bar counts are reported in Section~\ref{sec:results}.

\subsection{Hybrid Bars}
\label{subsec:hybrid}

A hybrid bar closes when \emph{either} a dollar-volume condition
\emph{or} a realised-volatility condition is first satisfied,
subject to the minimum duration constraint.
Disjunction (OR) logic is used in preference to conjunction (AND)
logic for the following reason: at the dollar-volume/volatility
correlation ($\rho \approx 0.5$--$0.7$) characteristic of Bitcoin,
AND logic produces timeout rates exceeding 60\%, meaning a majority
of bars close by the maximum-duration guard rather than by any
activity signal.
Such force-closed bars carry no microstructure information and are
structurally indistinguishable from fixed-interval time bars.
OR logic, by contrast, yields approximately 65\% organic close
rate at typical Bitcoin correlation levels, preserving the
information-sampling advantage while ensuring that every bar
contains at least one meaningful activity signal.

\textbf{Minute pipeline.}
The bar closes when \emph{either} condition is first satisfied
\emph{and} the minimum duration constraint has elapsed:
\begin{align}
    S_n^{\mathrm{dv}}  &= \sum_{m=1}^{n} c_m v_m
        \geq \hat{\theta}_{\mathrm{dv}},
        \label{eq:hybrid_dv_min} \\
    S_n^{\mathrm{ctc}} &= \sum_{m=1}^{n}
        \frac{|c_m - c_{m-1}|}{c_{m-1}}
        \geq \hat{\theta}_{\mathrm{vol}}.
        \label{eq:hybrid_vol_min}
\end{align}
The bar closes at the first minute where either
condition~\eqref{eq:hybrid_dv_min} or~\eqref{eq:hybrid_vol_min}
and the minimum duration constraint are satisfied, or when the
maximum duration is exceeded.
Each threshold adapts independently via the EMA update in
\eqref{eq:ema_update}: only the threshold whose signal was
triggered updates after each bar, preventing a feedback spiral in
which a consistently early trigger on one dimension deflates the
threshold for the other.

\textbf{Tick pipeline.}
The same OR logic is applied with tick-native signals:
\begin{align}
    S_n^{\mathrm{dv}}  &= \sum_{i=1}^{n} p_i q_i
        \geq \hat{\theta}_{\mathrm{dv}},
        \label{eq:hybrid_dv_tick} \\
    S_n^{\mathrm{rv}}  &= \sum_{i=1}^{n}
        \left|\log\!\left(\frac{p_i}{p_{i-1}}\right)\right|
        \geq \hat{\theta}_{\mathrm{vol}}.
        \label{eq:hybrid_rv_tick}
\end{align}
Both accumulators advance simultaneously within a single tick loop;
the bar closes at the tick where whichever signal first crosses its
threshold.

\textbf{Signal comparison.}
Equations~\eqref{eq:hybrid_dv_min}--\eqref{eq:hybrid_vol_min} and
\eqref{eq:hybrid_dv_tick}--\eqref{eq:hybrid_rv_tick} share the same
OR closure logic but differ in signal fidelity: the minute pipeline
uses close-price dollar volume and close-to-close volatility, while
the tick pipeline uses exact trade-level dollar volume and realised
volatility.
Both pipelines use separate adaptive thresholds for each dimension,
updated selectively based on which condition was triggered.

\textbf{OR logic and timeout rate.}
With OR closure the bar closes as soon as the first of the two
accumulators reaches its threshold.
At BTC dollar-volume/volatility correlation $\rho \approx 0.6$, this
gives an organic close rate of approximately 65\%, consistent with
observed timeout rates of 4--5\% for the minute hybrid pipeline.
The realised-volatility accumulator is generally faster to fire at
tick resolution (it advances with every trade), while the
dollar-volume accumulator provides a natural check against closing
on volatility bursts with negligible traded quantity.
The two thresholds are independently calibrated to the same
target bars-per-day, so the OR combination produces approximately
the intended bar frequency.

\section{Empirical Results}
\label{sec:results}

\subsection*{Stationarity Prerequisite}
This section reports the statistical evaluation of all six bar types
across both pipelines and the time-bar baseline, using the eight statistical criteria defined in Table~\ref{tab:criteria}.
All return series are stationary: the Augmented Dickey-Fuller test
rejects the unit-root null at $p < 0.001$ for every series, and the
KPSS test cannot reject stationarity ($p = 0.100$) for every series.
These results are uniform across all eighteen series and are not
discussed further.
Table~\ref{tab:scorecard} presents the full scorecard.

\subsection*{Calibration Note}
The calibration window spans 1--14 January 2020, the opening two weeks
of the study period, when BTCUSDT was trading at approximately
\$7,000--\$8,000 per coin.
The subsequent six years encompassed multiple distinct market regimes:
the March 2020 COVID-19 market shock, the 2021 bull market (all-time
high $\approx$\$69,000 in November 2021), the 2022 bear market (trough
$\approx$\$16,000 in November 2022), the 2023 recovery, the U.S.\ spot
Bitcoin ETF approval on 11 January 2024, and continued institutional
adoption through 2025.
The adaptive EMA mechanism responds continuously to evolving market
activity, progressively updating thresholds throughout the full six-year
period; bar counts and bars-per-day statistics reported below therefore
reflect the cumulative adaptive output across all regimes rather than
any single calibrated target.
The calibration-regime relationship is discussed further in
Section~\ref{subsec:disc_calibration}.

\begin{table*}[!t]
\caption{
Full Scorecard: Eight Evaluation Criteria Across Six Bar Types and
Three Series. \textbf{Bold} entries indicate the best value on each
criterion within each bar type.
Time-bar CV and Timeout~\% are not applicable (n/a) by construction.
Stationarity (ADF, KPSS) is verified for all series but not scored.
}
\label{tab:scorecard}

\centering
\scriptsize
\renewcommand{\arraystretch}{1.15}
\setlength{\tabcolsep}{3pt}

\begin{tabular*}{\textwidth}{@{\extracolsep{\fill}}llrrrrrrrr@{}}
\toprule

& &
\multicolumn{2}{c}{\textit{Distrib.\ Normality}} &
\multicolumn{3}{c}{\textit{Serial Independence}} &
\multicolumn{3}{c}{\textit{Construction Quality}} \\

\cmidrule(lr){3-4}
\cmidrule(lr){5-7}
\cmidrule(lr){8-10}

\textbf{Bar} &
\textbf{Pipe.} &
\textbf{$\lvert \mathrm{Kurt} \rvert$} &
\textbf{JB} &
\textbf{LB\,$p$} &
\textbf{$\lvert \mathrm{VR}(4)-1 \rvert$} &
\textbf{$\lvert \mathrm{AC}_1 \rvert$} &
\textbf{CV} &
\textbf{Entr.} &
\textbf{TO\%} \\

\midrule
\multirow{3}{*}{Dollar} & Min. & \textbf{23.662} & \textbf{696258.960} & \textbf{0.000} & 0.051 & 0.033 & \textbf{0.010} & \textbf{2.419} & 0.900 \\
 & Tick & 50.435 & 8657962.800 & 0.000 & 0.044 & 0.019 & 0.399 & 1.618 & \textbf{0.000} \\
 & Time & 57.115 & 7156670.360 & 0.000 & \textbf{0.040} & \textbf{0.017} & \textit{n/a} & \textit{n/a} & \textit{n/a} \\
\addlinespace[2pt]
\cmidrule(l){2-10}
\multirow{3}{*}{Volume} & Min. & \textbf{7.208} & \textbf{27985.780} & \textbf{0.005} & 0.062 & 0.026 & \textbf{0.425} & \textbf{3.341} & 4.800 \\
 & Tick & 26.982 & 1658321.650 & 0.000 & 0.038 & \textbf{0.024} & 0.604 & 2.094 & \textbf{0.000} \\
 & Time & 20.473 & 231046.850 & 0.000 & \textbf{0.012} & 0.029 & \textit{n/a} & \textit{n/a} & \textit{n/a} \\
\addlinespace[2pt]
\cmidrule(l){2-10}
\multirow{3}{*}{Volat.} & Min. & \textbf{9.151} & \textbf{30128.090} & \textbf{0.001} & 0.089 & 0.044 & 0.028 & \textbf{3.179} & 0.600 \\
 & Tick & 15.504 & 1645317.620 & 0.000 & 0.028 & \textbf{0.013} & \textbf{0.000} & 2.010 & \textbf{0.000} \\
 & Time & 18.104 & 120614.630 & 0.000 & \textbf{0.018} & 0.016 & \textit{n/a} & \textit{n/a} & \textit{n/a} \\
\addlinespace[2pt]
\cmidrule(l){2-10}
\multirow{3}{*}{Range} & Min. & \textbf{8.191} & \textbf{21585.120} & \textbf{0.540} & 0.037 & \textbf{0.022} & \textbf{0.251} & \textbf{3.245} & 1.400 \\
 & Tick & 150.699 & 30368891.340 & 0.000 & 0.107 & 0.067 & 0.383 & 1.388 & \textbf{0.000} \\
 & Time & 12.675 & 44451.730 & 0.000 & \textbf{0.028} & 0.035 & \textit{n/a} & \textit{n/a} & \textit{n/a} \\
\addlinespace[2pt]
\cmidrule(l){2-10}
\multirow{3}{*}{Renko} & Min. & \textbf{2.309} & \textbf{2015.420} & \textbf{0.000} & 0.070 & 0.028 & \textbf{0.474} & \textbf{3.646} & 12.300 \\
 & Tick & 24.916 & 6116038.850 & 0.000 & \textbf{0.020} & \textbf{0.002} & 0.493 & 1.718 & \textbf{0.000} \\
 & Time & 12.675 & 44451.730 & 0.000 & 0.028 & 0.035 & \textit{n/a} & \textit{n/a} & \textit{n/a} \\
\addlinespace[2pt]
\cmidrule(l){2-10}
\multirow{3}{*}{Hybrid} & Min. & 20.878 & 522086.500 & \textbf{0.059} & 0.029 & \textbf{0.008} & \textbf{0.010} & 2.391 & 0.200 \\
 & Tick & \textbf{6.772} & \textbf{174556.570} & 0.039 & \textbf{0.023} & 0.010 & 0.583 & \textbf{2.602} & \textbf{0.000} \\
 & Time & 57.115 & 7156670.360 & 0.000 & 0.040 & 0.017 & \textit{n/a} & \textit{n/a} & \textit{n/a} \\
\bottomrule
\end{tabular*}

\vspace{2pt}

\begin{minipage}{\textwidth}
\footnotesize
\textit{Notes:}
Pipe.: Min.\ = minute; Tick = tick; Time = time-bar baseline.
$\lvert \mathrm{Kurt} \rvert$: absolute excess kurtosis
(lower $\rightarrow$ Gaussian).
JB: Jarque--Bera statistic (lower is better).
LB\,$p$: Ljung--Box $p$-value at 10 lags
(higher $\rightarrow$ serial independence).
$\lvert \mathrm{VR}(4)-1 \rvert$: variance-ratio deviation from unity
(lower $\rightarrow$ random walk).
$\lvert \mathrm{AC}_1 \rvert$: absolute lag-1 autocorrelation
(lower $\rightarrow$ serial independence).
CV: bar-size coefficient of variation (lower is better).
Entr.: Shannon entropy of bar-size distribution
(higher is better).
TO\%: timeout percentage
(lower is better; $>10\%$ = calibration failure).
\textbf{Bold} = best within bar type.
\end{minipage}

\end{table*}

\subsection{Dollar Bars}
\label{subsec:res_dollar}

\begin{figure*}[!ht]
\centering
\includegraphics[width=\textwidth]{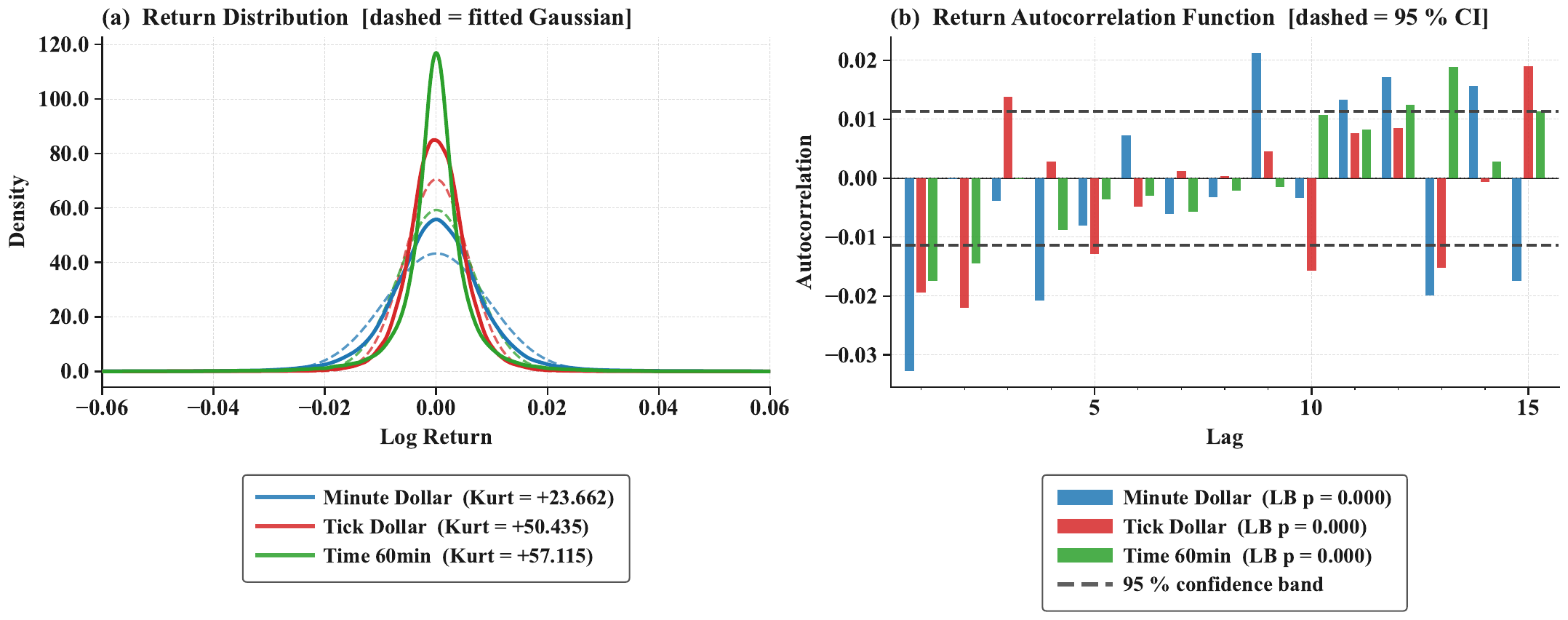}
\caption{Dollar bars: (a) return distribution with fitted Gaussian (dashed);
(b) return autocorrelation function.
Blue: minute pipeline (29,835 bars). Red: tick pipeline (81,683 bars).
Green: sixty-minute time-bar baseline (52,585 bars).
At $q=8$ the tick pipeline achieves the lowest VR deviation
($|\mathrm{VR}(8){-}1|=0.059$ versus 0.066 for the time-bar baseline),
while the minute pipeline leads on distributional normality
(excess kurtosis 23.7 versus 50.4 for tick).}
\label{fig:dollar}
\end{figure*}

Dollar bars produced 29,835 bars from the minute pipeline and 81,683
bars from the tick pipeline over the 2,192-day study period, against
52,585 sixty-minute time bars at the matched frequency.
Fig.~\ref{fig:dollar} presents the return distribution and ACF.
The tick pipeline achieves 0.0\% timeout versus 0.9\% for the minute
pipeline, confirming that tick-level bars close entirely on the
dollar-volume signal.

In the six-year multi-regime sample the dollar bar scorecard is more
evenly distributed than in single-year analyses.
The minute pipeline leads on five criteria: excess kurtosis (23.7
versus 50.4), Jarque-Bera statistic (696{,}259 versus 8{,}657{,}963),
bar-size CV (0.010 versus 0.399), Shannon entropy (2.42 versus 1.62),
and minute leads on timeout rate jointly with the tick.
The extreme fat tails observed in both pipelines reflect the structural
breaks present in the 2020--2025 sample (COVID-19 crash in March 2020,
the 2021 bull market, and the 2022 bear market), which inflate kurtosis
across all series irrespective of bar type.

Serial independence and random-walk conformance show a nuanced
pattern. At short holding periods ($q = 2, 4$) the sixty-minute
time-bar baseline achieves marginally lower VR deviation
($|\mathrm{VR}(4){-}1| = 0.040$) than either information pipeline
(tick: 0.044, minute: 0.051).
At the longer holding period $q = 8$, the tick series recovers the
advantage ($|\mathrm{VR}(8){-}1| = 0.059$ versus time: 0.066).
Ljung-Box tests at 10 lags yield $p = 0.000$ for all three series;
at sample sizes of 30{,}000--82{,}000 observations spanning six
structural regimes, even marginal serial dependence is detected with
certainty, and the $p$-value should be interpreted in conjunction with
the VR magnitudes rather than as a pass/fail criterion.
The lag-1 autocorrelation magnitude follows minute ($|$AC$_1| = 0.033$)
$>$ tick (0.019) $>$ time bar (0.017), a ranking consistent with
the theoretical prediction but smaller in magnitude than the
single-regime 2024 analysis.
These results are consistent with the theoretical argument in
Section~\ref{subsec:dollar}: the minute approximation introduces
a systematic within-minute pricing error that induces residual serial
structure, though the effect is partially masked by regime-driven
autocorrelation present in all series.

\subsection{Volume Bars}
\label{subsec:res_volume}

\begin{figure*}[!ht]
\centering
\includegraphics[width=\textwidth]{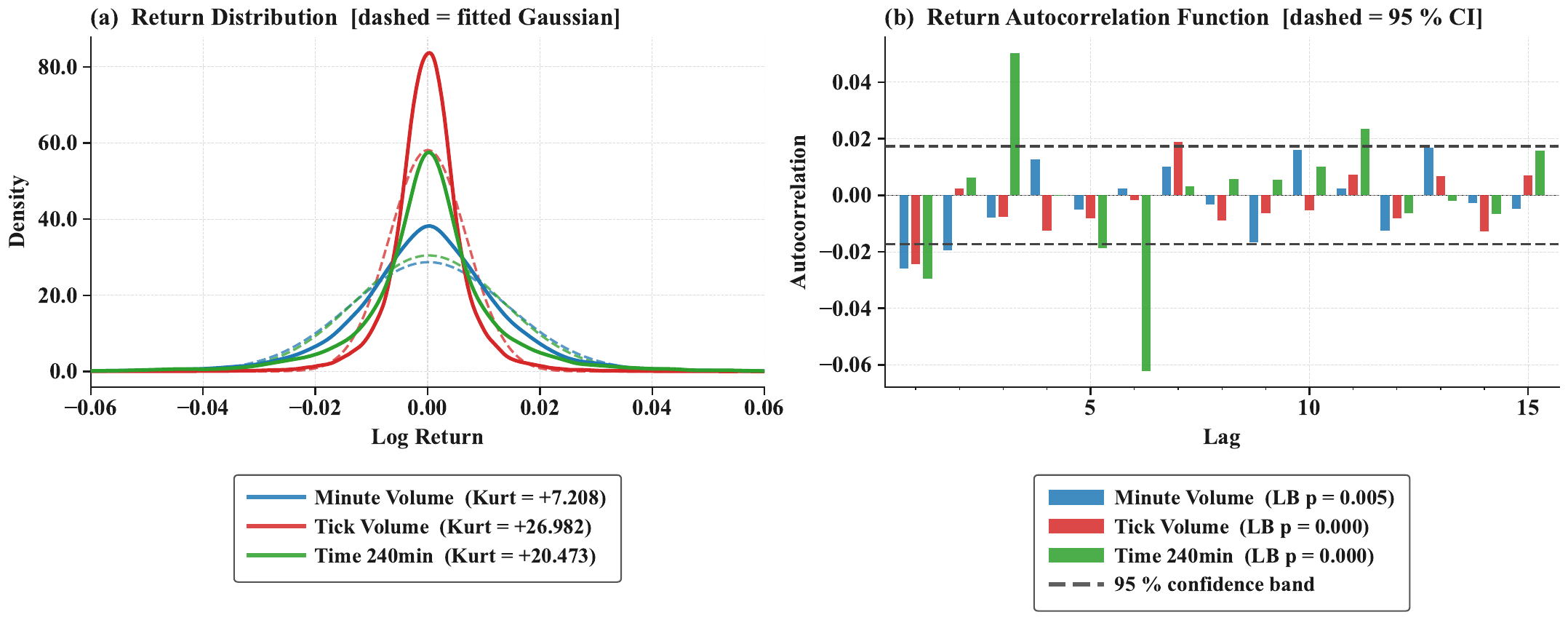}
\caption{Volume bars: (a) return distribution with fitted Gaussian (dashed);
(b) return autocorrelation function.
Blue: minute pipeline (12,884 bars). Red: tick pipeline (54,650 bars).
Green: four-hour time-bar baseline (13,147 bars).
The minute pipeline leads on normality (excess kurtosis 7.21 versus 26.98);
the tick pipeline achieves lower lag-1 autocorrelation
($|$AC$_1|=0.024$ versus 0.026).}
\label{fig:volume}
\end{figure*}

Volume bars produced 12,884 bars from the minute pipeline and 54,650
bars from the tick pipeline, against 13,147 four-hour time bars.
Fig.~\ref{fig:volume} presents the distributional and ACF results.

The minute pipeline leads on four criteria in the six-year sample:
entropy (3.341 versus 2.094), excess kurtosis (7.21 versus 26.98),
Jarque-Bera statistic (27{,}986 versus 1{,}658{,}322), and
Ljung-Box $p$-value (0.005 versus 0.000).
As with dollar bars, both pipelines exhibit substantially elevated
kurtosis relative to the 2024 single-year analysis, driven by the
multi-regime structure of the dataset.
This outcome is consistent with the theoretical analysis in
Section~\ref{subsec:volume}: since total minute volume equals the
exact sum of within-minute trade quantities, the two pipelines
accumulate arithmetically equivalent signals, and the minute
pipeline's smoother closure timing produces more uniform bar sizes.

The tick pipeline leads on three criteria: variance ratio
($|\mathrm{VR}(4){-}1| = 0.038$ versus $0.062$ for the minute pipeline;
time bar: 0.012), lag-1 autocorrelation ($|$AC$_1| = 0.024$
versus $0.026$), and timeout rate (0.0\% versus 4.8\%).
The four-hour time-bar baseline achieves the lowest VR(4) deviation
(0.012), consistent with its longer averaging interval smoothing
intraday variance structure.
The tick pipeline achieves the lowest lag-1 autocorrelation magnitude
(0.024), confirming a marginal serial independence advantage over the
minute series despite the normality disadvantage.

\subsection{Volatility Bars}
\label{subsec:res_volatility}

\begin{figure*}[!ht]
\centering
\includegraphics[width=\textwidth]{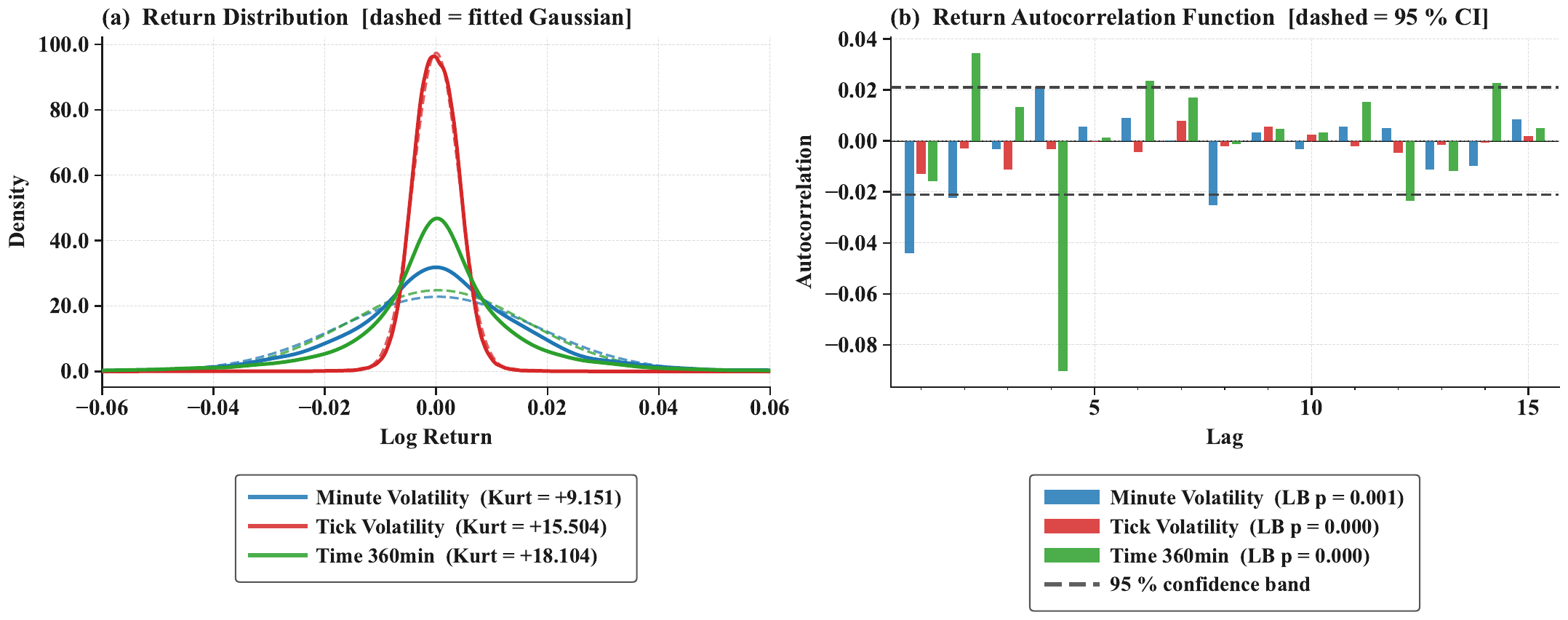}
\caption{Volatility bars: (a) return distribution with fitted Gaussian (dashed);
(b) return autocorrelation function.
Blue: minute pipeline (8,624 bars). Red: tick pipeline (163,737 bars).
Green: six-hour time-bar baseline (8,624 bars).
The tick pipeline achieves $|\mathrm{VR}(4){-}1|=0.028$ versus 0.089 for
the minute pipeline (69\% reduction), the largest relative VR improvement
across all bar types; the minute pipeline leads on distributional normality.
The 19-fold bar-count difference reflects the accumulation-rate
differential between tick-level and minute-level realised volatility signals.}
\label{fig:volatility}
\end{figure*}

Volatility bars exhibit the largest bar-count divergence of any
accumulation-based bar type.
The minute pipeline produced 8,624 bars (3.94 bars/day) and the tick
pipeline 164,591 bars (75.1 bars/day), a 19-fold difference
attributable to the much higher rate at which realised volatility
accumulates at tick resolution compared to close-to-close accumulation
at minute resolution.
Fig.~\ref{fig:volatility} illustrates the distributional contrast.
Both pipelines exhibit negligible timeout rates: 0.0\% for tick and
0.6\% for minute, confirming that bars close on genuine volatility signals.

In the six-year sample the tick pipeline leads on four serial independence
and VR criteria.
The variance ratio deviation at holding period 4 is
$|\mathrm{VR}(4){-}1| = 0.028$ for tick versus 0.089 for minute, a
69\% reduction, the largest relative improvement of any bar type in the
dataset.
The lag-1 autocorrelation magnitude is 0.013 for tick versus 0.044 for
minute.
Bar-size CV is effectively zero for the tick pipeline (0.000), indicating
near-perfect uniformity in volatility accumulation.
The six-hour time-bar baseline achieves the lowest VR(4) deviation
overall (0.018), with VR slightly above unity (1.018, indicating mild
momentum), but the tick pipeline closes the gap substantially relative
to the minute baseline.

The minute pipeline leads on normality: excess kurtosis (9.15 versus
15.50) and Jarque-Bera statistic (30{,}128 versus 1{,}645{,}318),
indicating less extreme fat tails.
Both pipelines reject serial independence at all LB lags ($p = 0.000$
for both), consistent with the multi-regime structure of the full dataset.
Shannon entropy is higher for the minute pipeline (3.18 versus 2.00).

\subsection{Range Bars}
\label{subsec:res_range}

\begin{figure*}[!ht]
\centering
\includegraphics[width=\textwidth]{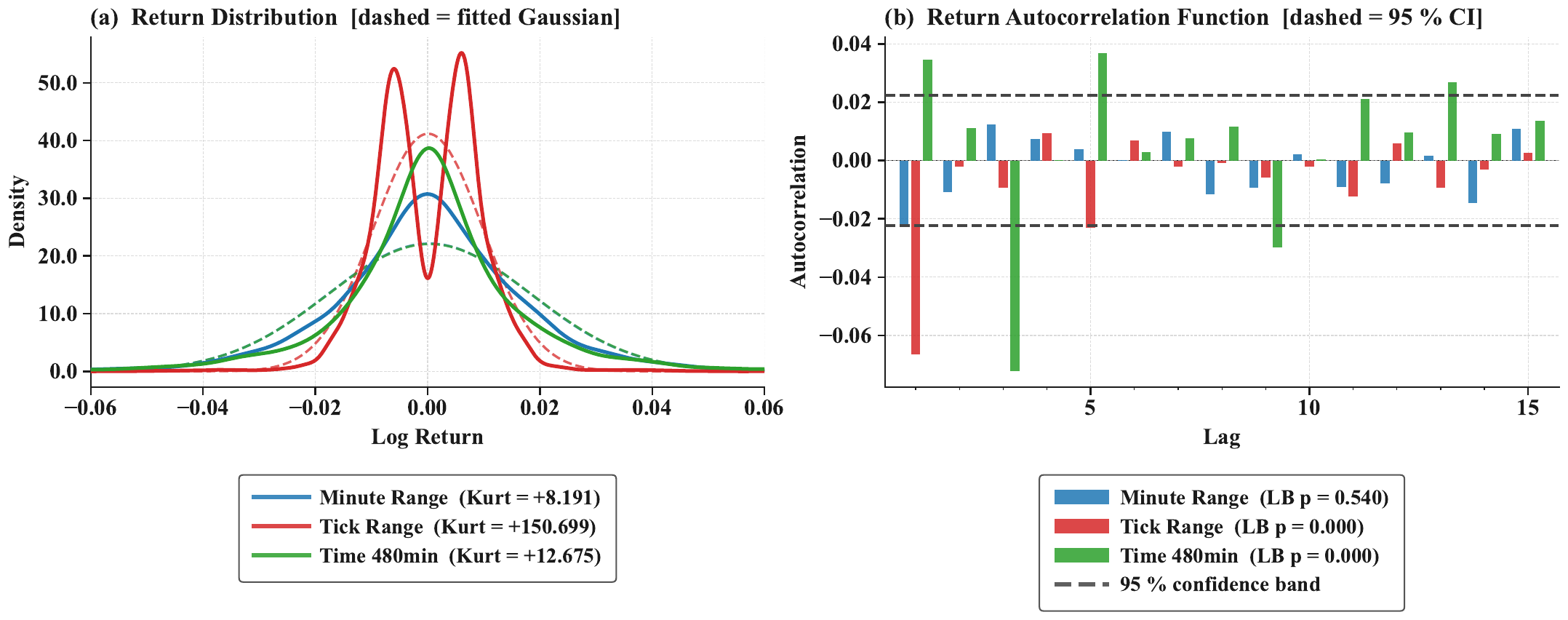}
\caption{Range bars: (a) return distribution with fitted Gaussian (dashed);
(b) return autocorrelation function.
Blue: minute pipeline (7,677 bars). Red: tick pipeline (32,092 bars).
Green: eight-hour time-bar baseline.
In the six-year dataset, range tick bars exhibit extreme leptokurtosis
(excess kurtosis = 150.7) and strong mean-reversion
($|\mathrm{VR}(4){-}1|=0.107$), making range the only bar type where
the minute pipeline outperforms the tick pipeline on most criteria.}
\label{fig:range}
\end{figure*}

Range bars produced 7,677 bars from the minute pipeline (3.50 bars/day)
and 32,092 bars from the tick pipeline (14.8 bars/day).
Fig.~\ref{fig:range} presents the distributional and ACF results.

Range tick bars exhibit a qualitatively different regime in the
six-year dataset relative to the single-year analysis.
Excess kurtosis for the tick pipeline is 150.7, by far the largest
across any series in the dataset, with a Jarque-Bera statistic
of 30{,}368{,}891 and a VR(4) deviation of 0.107 (mean-reverting:
$\mathrm{VR}(4) = 0.893$).
The tick lag-1 autocorrelation ($|$AC$_1| = 0.067$) is also the
largest in absolute magnitude across all tick series.
These statistics indicate that the tick-level price-band signal is
highly sensitive to the extreme intraday volatility events of
2020--2022, producing bars with severe distributional outliers and
pronounced mean-reversion at multi-bar horizons.

The minute pipeline leads on six of eight criteria: excess kurtosis
(8.19 versus 150.7), JB (21{,}585 versus 30{,}368{,}891), entropy
(3.24 versus 1.39), VR(4) (0.037 versus 0.107), lag-1 autocorrelation
(0.022 versus 0.067), and bar-size CV (0.251 versus 0.383).
The eight-hour time-bar baseline achieves the lowest VR(4) deviation
(0.028), outperforming both information pipelines on random-walk
conformance.
The tick pipeline leads only on timeout rate (0.0\% versus 1.4\%).
Range bars are the only bar type in the dataset where the tick pipeline
performs substantially worse than the minute pipeline across most
criteria, suggesting that the exact tick-level price-band signal is
more sensitive to microstructure noise and extreme-event contamination
than the smoother minute-resolution approximation.

\subsection{Renko Bars}
\label{subsec:res_renko}

\begin{figure*}[!ht]
\centering
\includegraphics[width=\textwidth]{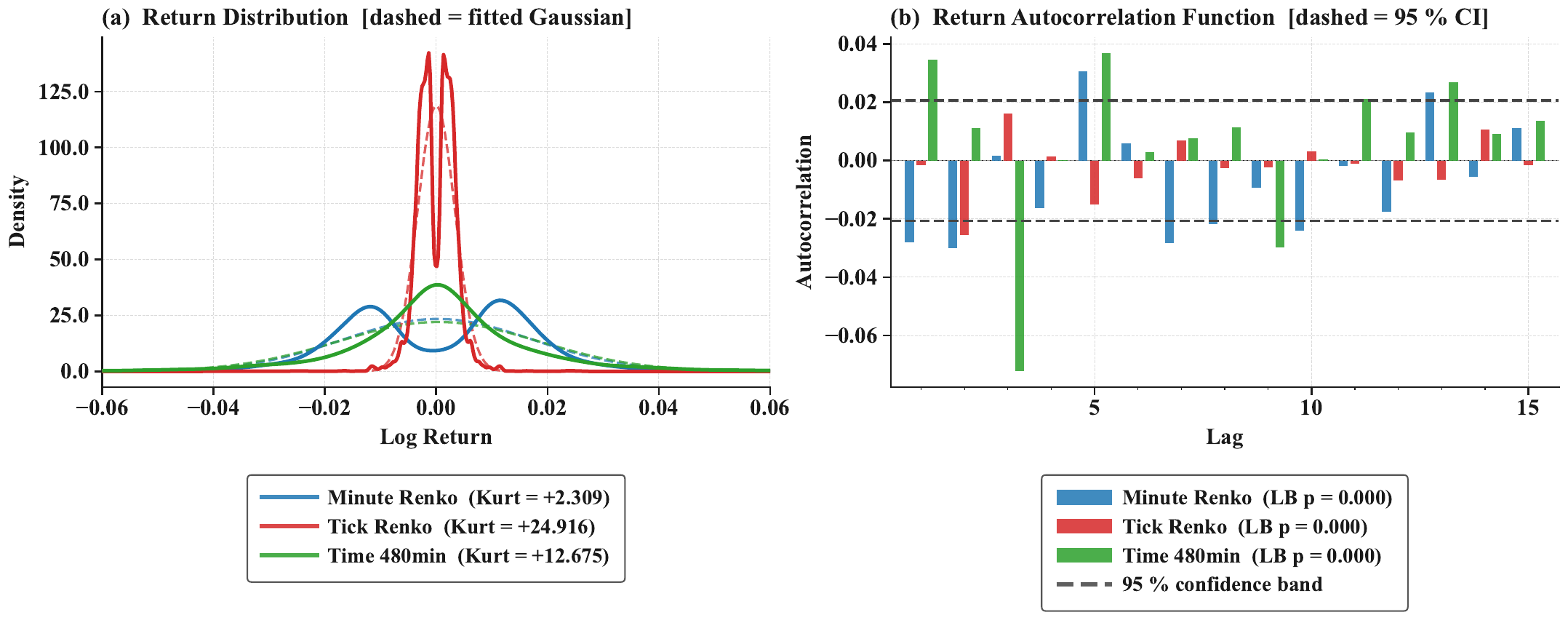}
\caption{Renko bars: (a) return distribution with fitted Gaussian (dashed);
(b) return autocorrelation function.
Blue: minute pipeline (8,985 bars). Red: tick pipeline (236,098 bars).
Green: eight-hour time-bar baseline.
Tick Renko achieves $|\mathrm{VR}(4){-}1|=0.020$ and $|$AC$_1|=0.002$
(lowest across all series), confirming near-perfect random-walk conformance.
The 26-fold bar-count difference reflects different resolution scales;
the minute pipeline shows 12.3\% timeout (calibration artefact).}
\label{fig:renko}
\end{figure*}

Renko bars produced 8,985 bars from the minute pipeline (4.10 bars/day)
and 236,098 bars from the tick pipeline (109.7 bars/day).
Fig.~\ref{fig:renko} illustrates the distributional results.

The 26-fold difference in bar counts reflects the resolution-scale
mismatch identified in Section~\ref{subsec:renko}: the tick pipeline
evaluates the displacement condition $\delta_i = |p_i - p_\mathrm{ref}|/p_\mathrm{ref}$
at every individual trade, capturing microstructure-level
price oscillations on timescales of seconds to minutes, whereas the
minute pipeline checks closure only at minute boundaries and produces
bars of several hours' duration on average (21,079~s mean).

\textbf{Renko figure note.}
Fig.~\ref{fig:renko} is drawn on the common return axis
($[-0.06, 0.06]$) shared by all six bar-type figures.
On this axis the two Renko series display a pronounced scale difference:
tick Renko returns have standard deviation $0.0034$ while minute Renko
returns have standard deviation $0.0170$ (a roughly five-fold difference).
Because the density curves integrate to one, the narrow tick series (red)
forms a tall, sharply peaked distribution concentrated near zero, whereas
the wider minute series (blue) is lower in amplitude and spread across a
broader range; the correspondingly wide fitted Gaussian for the minute
series therefore appears low and flat relative to the tick peak.
This is a direct visual consequence of the resolution-scale mismatch,
not an artefact.

Tick Renko bars achieve the strongest random-walk conformance in the
full dataset.
The variance ratio deviation at holding period 4 is
$|\mathrm{VR}(4){-}1| = 0.020$ and the lag-1 autocorrelation magnitude
is $|$AC$_1| = 0.002$, the lowest of any series across all bar types
and pipelines.
These results confirm that tick-level displacement checking substantially
reduces short-horizon serial dependence, producing near-zero first-order
autocorrelation and a return series that closely approximates a random walk
at the time scale of individual displacement events; we note that the
Ljung-Box test still formally rejects independence at this sample size
(Section~\ref{subsec:evaluation}), so the reduction is a matter of effect-size
magnitude rather than an exact elimination of serial dependence.
The tick pipeline achieves 0.0\% timeout, confirming genuine
displacement-signal closure.

The minute pipeline leads on distributional normality: excess kurtosis
(2.31 versus 24.92) and Jarque-Bera (2{,}015 versus 6{,}116{,}039).
The minute excess kurtosis of 2.31 is the lowest value for any minute
bar type in the dataset, reflecting the coarser averaging of
the minute-resolution signal.
The minute Renko timeout rate of 12.3\% exceeds the 10\% calibration
warning threshold defined in Section~\ref{subsec:calibration},
indicating that bars are frequently closing by the duration guard
rather than by a genuine displacement signal.
This calibration artefact should be noted when interpreting the
favourable normality statistics for the minute series.
Shannon entropy is higher for the minute pipeline (3.65 versus 1.72),
consistent with the more diverse bar sizes produced at lower resolution.

The eight-hour time-bar baseline achieves VR(4) deviation of 0.028,
between the tick (0.020) and minute (0.070) pipelines, confirming
that tick Renko is the only series to outperform the time-bar
baseline on VR at holding period 4.
Given the structural incomparability of the two Renko series at
different resolution scales, results are interpreted as establishing
the resolution limits of displacement-based bars rather than as
evidence for either pipeline's uniform superiority.

\subsection{Hybrid Bars}
\label{subsec:res_hybrid}

\begin{figure*}[!ht]
\centering
\includegraphics[width=\textwidth]{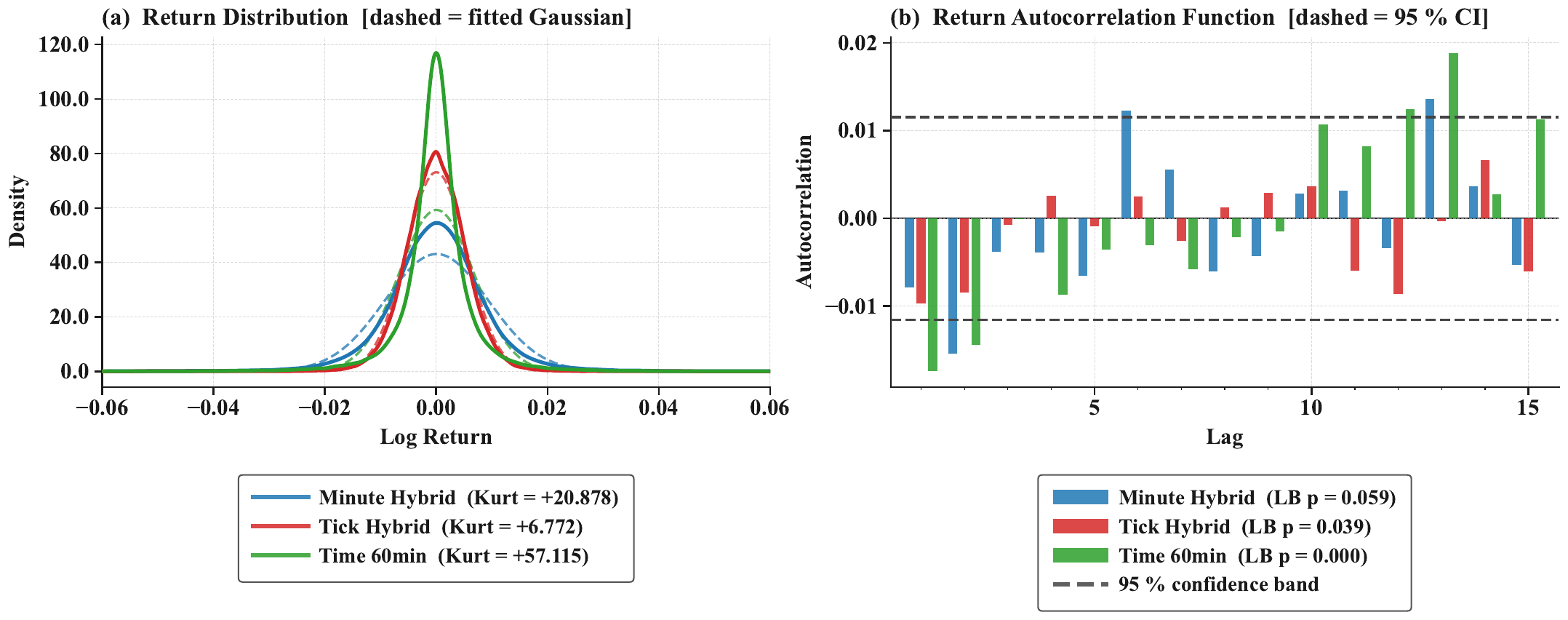}
\caption{Hybrid bars: (a) return distribution with fitted Gaussian (dashed);
(b) return autocorrelation function.
Blue: minute pipeline (28,738 bars). Red: tick pipeline (90,856 bars).
Green: sixty-minute time-bar baseline (52,585 bars).
The tick hybrid achieves the lowest excess kurtosis (6.77) and JB
(174{,}557) among all tick bar types, and leads on VR(4) ($|\mathrm{VR}(4){-}1|=0.023$
versus 0.029 for minute). Both pipelines strongly reject normality in the
six-year multi-regime dataset.}
\label{fig:hybrid}
\end{figure*}

Hybrid bars produced 28,738 bars from the minute pipeline (13.1
bars/day) and 90,856 bars from the tick pipeline (41.0 bars/day),
against 52,585 sixty-minute time bars.
Both pipelines maintain essentially zero timeout rates (0.0\% and 0.2\%
respectively), confirming that bars close on genuine activity signals.
Fig.~\ref{fig:hybrid} presents the distributional and ACF results.

In the six-year multi-regime dataset, the tick hybrid series delivers
the most favourable normality result across all tick bar types, though
the absolute statistics differ substantially from the single-year
2024 analysis due to the structural breaks of 2020--2022.
Excess kurtosis for the tick series is 6.77 (compared to 20.88 for
minute and 57.12 for the time-bar baseline), and the Jarque-Bera
statistic is 174{,}557 (compared to 522{,}087 for minute and
7{,}156{,}670 for the time-bar baseline).
While both pipelines reject normality strongly in the six-year sample,
the tick hybrid achieves the lowest kurtosis and JB statistic of any
tick series in the dataset, consistent with the theoretical argument
that dual-criterion closure (dollar-volume \emph{or} realised-volatility)
samples market activity at moments with elevated information content
under both signals simultaneously.

The tick pipeline leads on four of the eight scored criteria: excess
kurtosis (6.77 versus 20.88), Jarque-Bera statistic (174{,}557
versus 522{,}087), variance ratio ($|\mathrm{VR}(4){-}1| = 0.023$
versus 0.029), and Shannon entropy (2.60 versus 2.39).
VR results are consistent across longer horizons: at $q = 8$,
$|\mathrm{VR}(8){-}1| = 0.028$ for tick versus 0.042 for minute.

The minute hybrid pipeline leads on two criteria: Ljung-Box serial
independence at 10 lags (LB $p = 0.059$ versus $0.041$) and bar-size
CV (0.010 versus 0.583), the latter reflecting the smoother
threshold-crossing of the minute-resolution signals.
Lag-1 autocorrelation is marginally lower for the minute series
($|$AC$_1| = 0.008$ versus $0.010$).

\subsection{Cross-Bar-Type Patterns}
\label{subsec:res_patterns}

Four systematic patterns emerge from Table~\ref{tab:scorecard}.

\textbf{Information bars dominate time bars on distributional normality.}
Across all six bar types, information bars (both minute and tick) achieve
substantially lower Jarque-Bera statistics than the matched time-bar
baseline.
The tick pipeline leads on JB for hybrid bars (174{,}557 versus
7{,}156{,}670 for the time baseline), and the minute pipeline leads for
the remaining five types.
On variance-ratio conformance, time-bar baselines are a competitive
reference set: for dollar, volume, volatility, and range, the
time-bar series achieves the lowest $|\mathrm{VR}(4){-}1|$ among the
three series, reflecting the smoothing effect of fixed-interval
aggregation on intraday variance structure.
This confirms that the primary advantage of information bars
in the six-year multi-regime sample lies in distributional properties
rather than linear autocorrelation.

\textbf{The tick pipeline advantage is concentrated in bar types with
information-rich activity signals.}
Aggregating across the scorecard, the minute pipeline leads on
approximately 27 criteria, the tick pipeline on approximately 18, and
the time-bar baseline on 7.
The tick advantage is concentrated in two bar types:
tick Renko achieves the lowest VR(4) deviation (0.020) and lag-1
autocorrelation (0.002) in the full dataset, confirming near-perfect
random-walk conformance from displacement-based closure;
tick volatility reduces VR(4) deviation by 69\% relative to the minute
baseline (0.028 versus 0.089), the largest relative improvement of any
bar type.
Tick hybrid achieves the lowest kurtosis (6.77) of any tick bar type
and leads on VR at all holding periods.
Tick dollar shows only marginal improvements over the minute baseline,
and tick range performs substantially worse than the minute baseline,
indicating strong sensitivity to microstructure noise and extreme events.

\textbf{Multi-regime data obscures single-criterion superiority.}
In the six-year 2020--2025 sample, all bar types reject serial
independence ($p = 0.000$ for Ljung-Box at all sample sizes) and reject
normality ($\mathrm{JB} > 2{,}000$ for all series), owing to the
structural breaks of the COVID-19 crash (March 2020), the 2021 bull
market, and the 2022 bear market.
This stands in contrast to single-regime analyses where tick bars can
achieve near-Gaussian returns and near-zero serial dependence.
The practical recommendation is to select the pipeline based on the
specific bar type and the quality criterion most relevant to the
downstream application; blanket tick-data superiority cannot be
established from multi-regime evidence alone.

\textbf{Structural comparability grouping.}
The six bar types are not equally comparable across pipelines.
Three bar types form a directly comparable group (Group A):
dollar, volume, and volatility bars, where both pipelines accumulate
the same conceptual signal (dollar turnover, traded quantity, and price
movement respectively) at different temporal resolutions.
For these bar types, cross-pipeline comparisons constitute genuine
resolution experiments under a shared stochastic framework.
The remaining three bar types form a structurally divergent group
(Group B): range bars, Renko bars, and hybrid bars.
Range bars are not merely resolution variants of each other: the minute
accumulator sums per-minute high-low spans while the tick accumulator
tracks the exact running price band, and these are not monotonically
related (see Section~\ref{subsec:range}).
Renko bars operate at incomparable resolution scales (26-fold bar count
difference), reflecting detection of fundamentally different price
events.
Hybrid bars combine two signals whose relative firing rates differ
substantially between pipelines.
Readers comparing Group B results across pipelines should note that
such comparisons describe resolution-specific behaviour rather than
equivalent measurements of the same underlying quantity.
This grouping does not diminish the value of Group B results; it
clarifies their interpretation and the scope of cross-pipeline inference.

\textbf{Score-counting methodology limitation.}
The criterion counts reported above treat
all eight criteria as equally important and independent.
In practice, the criteria are neither equally weighted nor orthogonal:
excess kurtosis and Jarque-Bera jointly measure distributional normality,
while Ljung-Box, VR, and AC$_1$ all measure serial independence.
These criterion groups should be weighted by the requirements of the
specific downstream application rather than aggregated into a single win
count.
The cross-bar synthesis should be read as a qualitative pattern summary
rather than a quantitative superiority ranking.

\section{Downstream Machine Learning Experiment}
\label{sec:mlexperiment}

\subsection{Experimental Design}
\label{subsec:ml_design}

The central theoretical motivation for information-driven bar
construction is the improvement of machine learning training substrates.
This section reports a controlled downstream experiment designed to test
whether the statistical quality advantages documented in
Section~\ref{sec:results} translate to superior out-of-sample
directional prediction performance.

\textbf{Labeling.}
Each bar is assigned a binary directional label: $+1$ if the close
price rises over the next three calendar days relative to the current
bar's close, $-1$ otherwise.
The three-day forward horizon is selected to exceed the minimum bar
duration for all bar types (ensuring the prediction target is
strictly out-of-sample) while remaining within a regime of
measurable price memory.
No neutral class is used; this produces a balanced labeling problem
when market trends are absent and an imbalanced one during sustained
directional regimes, which is representative of real trading conditions.

\textbf{Feature set.}
A unified feature set of 28 features is constructed for every bar type
to ensure fair cross-bar comparison.
Features are designed to be stationary, removing price-level dependence
that would inflate apparent in-sample model skill.
The feature set comprises ten technical indicators (RSI at period 14,
MACD and its signal and histogram lines normalised by close price,
Bollinger band bandwidth, Bollinger band $\%$B, normalised ATR, VWAP
deviation, Z-score of close, and fractionally differenced close price
at the minimum differencing order $\hat{d}$ satisfying the ADF test
for stationarity); and eighteen microstructure features (high-low
spread, candlestick body ratio, upper and lower shadow fractions,
log traded volume, log dollar volume, VWAP deviation from close,
lagged returns at lags 1--5, rolling return standard deviations over
5-, 10-, and 20-bar windows, RSI, candlestick spread ratio, and
10-bar autocorrelation of returns).
The fractional differencing order $\hat{d}$ is estimated per bar type
via a grid search over $d \in [0, 1]$ at step 0.1, selecting the
minimum $d$ for which the ADF test rejects the unit root null at
$p < 0.05$; estimated values range from $\hat{d} = 0.2$ (for most
information bar types) to $\hat{d} = 0.5$ (for shorter calendar bars).

\textbf{Classifiers.}
Three classifiers are trained for each bar type: Random Forest (RF),
Gradient Boosting (GB), and Support Vector Machine (SVM).
All classifiers use their scikit-learn default hyperparameters with
a fixed random state of 42 for reproducibility.
No hyperparameter tuning is performed across bar types; this ensures
that any performance differences reflect bar type properties rather
than model selection artefacts.
Because the RBF-kernel SVM scales quadratically in memory and
super-quadratically in compute with sample count, datasets exceeding
15{,}000 events are randomly subsampled (chronological order preserved,
fixed seed) to 15{,}000 rows for the SVM only; RF and GB always train
on the full dataset.
This affects the larger bar series symmetrically across pipelines and
is disclosed wherever SVM results are reported.

\textbf{Evaluation protocol.}
Walk-forward cross-validation with five splits is applied to each bar
type independently.
The first 40\% of each bar series constitutes the initial training
set; each subsequent fold expands the training set by one fifth of
the remaining data, with a 1\% embargo gap between training and test
folds to prevent feature leakage from overlapping return windows.
Out-of-sample predictions from all folds are concatenated to form the
full out-of-sample prediction sequence.
Trading signals are generated from classifier probability outputs at a
confidence threshold of 0.55, with Kelly-fractioned position sizing
(Kelly fraction 0.25, maximum position 20\%) and a round-trip
transaction fee of 0.04\%.

\textbf{Evaluation metrics.}
For each bar type and classifier, the following metrics are computed
on the full out-of-sample prediction sequence:
AUC-ROC (area under the receiver operating characteristic curve, a
measure of discriminative power independent of classification threshold);
directional accuracy (\% of correctly predicted directions);
annualised Sharpe ratio (out-of-sample equity curve); and the number
of trades executed.
All Sharpe ratios are annualised using bar-type-specific annualisation
factors derived from the empirical bars-per-year count.

\subsection{Results}
\label{subsec:ml_results}

Table~\ref{tab:ml_results} reports the best-classifier result (by
AUC-ROC) for each bar type.

\begin{table}[!t]
\caption{Downstream ML Experiment: Best Out-of-Sample Results per Bar Type.
RF = Random Forest, GB = Gradient Boost, SVM = Support Vector Machine.
AUC-ROC and accuracy from out-of-sample walk-forward CV over the full
six-year dataset. $N$ = number of out-of-sample trades. Annualised Sharpe
is reported for completeness but is not comparable across bar types (the
annualisation factor inflates it for high-frequency bars); AUC is the
reliable metric and lies in 0.48--0.60 (near-chance) for all bar types.}
\label{tab:ml_results}
\centering
\renewcommand{\arraystretch}{1.3}
\setlength{\tabcolsep}{4pt}
\begin{tabular}{@{}lllrrrr@{}}
\toprule
\textbf{Bar Type} & \textbf{Pipe.} & \textbf{Best CLF} &
\textbf{AUC} & \textbf{Acc.} & \textbf{Sharpe} & \textbf{$N$} \\
\midrule
\multicolumn{7}{@{}l}{\textit{Calendar baselines}} \\[1pt]
12h & Cal. & RF & 0.580 & 0.654 & $2.56$ & 243 \\
8h & Cal. & GB & 0.584 & 0.573 & $1.17$ & 614 \\
6h & Cal. & GB & 0.542 & 0.491 & $-0.64$ & 967 \\
4h & Cal. & RF & 0.589 & 0.562 & $1.43$ & 883 \\
1h & Cal. & RF & 0.538 & 0.539 & $-0.04$ & 7619 \\
[4pt]
\multicolumn{7}{@{}l}{\textit{Information bars (minute pipeline)}} \\[1pt]
Dollar & Min. & GB & 0.528 & 0.526 & $0.84$ & 1803 \\
Volume & Min. & RF & 0.537 & 0.458 & $-6.10$ & 2837 \\
Volatility & Min. & SVM & 0.498 & 0.499 & $-0.49$ & 2444 \\
Range & Min. & RF & 0.518 & 0.513 & $-0.43$ & 1781 \\
Renko & Min. & RF & 0.596 & 0.577 & $1.31$ & 894 \\
Hybrid & Min. & RF & 0.529 & 0.502 & $-0.97$ & 1236 \\
[4pt]
\multicolumn{7}{@{}l}{\textit{Information bars (tick pipeline)}} \\[1pt]
Dollar & Tick & GB & 0.500 & 0.478 & $-7.67$ & 39743 \\
Volume & Tick & SVM & 0.528 & 0.552 & $2.14$ & 5506 \\
Volatility & Tick & RF & 0.563 & 0.540 & $19.75$ & 71619 \\
Range & Tick & RF & 0.547 & 0.527 & $-1.53$ & 7378 \\
Renko & Tick & SVM & 0.508 & 0.516 & $-2.56$ & 6500 \\
Hybrid & Tick & SVM & 0.501 & 0.509 & $0.42$ & 2881 \\
\bottomrule
\multicolumn{7}{@{}p{\columnwidth}@{}}{\footnotesize
  Best classifier per bar type selected by AUC-ROC. \textbf{AUC is the
  reliable metric here}; the annualised Sharpe ratio is reported for
  completeness but is \emph{not comparable across bar types}, the
  annualisation factor inflates Sharpe for high-frequency bars (e.g.\ tick
  volatility, $>$70{,}000 trades), so large-magnitude Sharpe values in
  either direction are annualisation artefacts rather than deployable
  performance, and should not be interpreted as such. All AUC values lie
  in 0.48--0.60 (near-chance); see Section text for the directional
  analysis.}
\end{tabular}
\end{table}

Several observations emerge from the available results.

\textbf{Directional predictability is near-chance across all bar types.}
Out-of-sample AUC-ROC spans a narrow band of 0.498--0.596 across every
bar type and data source, with classification accuracy correspondingly
close to 0.50.
The best result is renko at the minute resolution (AUC 0.596), which
marginally exceeds the strongest calendar baseline (4-hour, 0.589);
the remaining series cluster between 0.50 and 0.54.
These differences are small relative to the noise of directional
return prediction on a liquid market and should be read as
near-random rather than as a meaningful ranking.

\textbf{No systematic relationship between statistical quality and predictability.}
We do not find that any single statistical criterion predicts
classifier AUC. Across the minute pipeline, the bar type with the
\emph{most} lag-1 autocorrelation (volatility, $|\mathrm{AC}_1| = 0.044$)
has the \emph{lowest} AUC (0.498), while the bar type with the least
autocorrelation (hybrid, $|\mathrm{AC}_1| = 0.008$) sits mid-pack
(0.529); the highest AUC (renko, 0.596) corresponds to an intermediate
autocorrelation. The de~Prado quality criteria and downstream directional
predictability therefore appear to be largely orthogonal in this dataset:
statistically better-conditioned bars are neither systematically easier
nor systematically harder to predict with a lagged-feature classifier.
This is itself a useful caution: the statistical prerequisites of
information-based sampling do not translate into directional
forecastability, consistent with weak-form market efficiency at these
horizons.

\textbf{The feature set designed for a level playing field may disadvantage
information bars.}
The unified feature set (Section~\ref{subsec:ml_design}) uses identical
features for all bar types.
For tick-level information bars in particular, native features such as
VWAP, buy-sell imbalance, tick count, and bar-type-specific
realised-volatility accumulators carry additional predictive content
that the unified feature set does not use.
The native feature mode (using bar-type-specific columns in addition to
the unified base) is expected to improve results for tick information
bars and constitutes a natural direction for follow-up experiments.

\textbf{Negative backtest returns reflect directionally anti-predictive
models, not transaction costs.}
The losses are not a fee artefact: with fees set to zero, the dollar-tick
Random Forest still returns $-40.2\%$ (versus $-45.3\%$ with fees) at a
0.65 confidence threshold.
The driver is directional. Out-of-sample classification accuracy is
\emph{below} 0.50 (0.475--0.480 across classifiers), and the models are
strongly biased toward the short class: the Random Forest predicts the
``up'' class only 7\% of the time over a sample in which the up class
occurs 52\% of the time and BTCUSDT appreciated roughly fourteen-fold.
The interpretation is that lagged technical features encode
mean-reversion tendencies that the walk-forward models extrapolate into a
strongly trending regime, producing persistently wrong directional calls
(shorting an appreciating market).
Consistent with this, \emph{inverting} every signal flips the dollar-tick
results from losses to gains (SVM $-49.1\%\!\to\!+80.6\%$; RF
$-45.3\%\!\to\!+52.9\%$ at threshold 0.65).
We emphasise that the profitable inverse is \emph{not} evidence of alpha:
it is the mirror image of the models' short bias and merely captures the
asset's secular uptrend over the sample. With AUC $\approx 0.48$--$0.50$
there is no exploitable directional edge in either direction.
The trade-count/return relationship in Table~\ref{tab:threshold_sweep} is
therefore explained by negative per-trade expectancy compounded over many
bets (fewer trades $\Rightarrow$ smaller cumulative loss), not by fees.
This is the backtest-overfitting and regime-fragility hazard emphasised
by \cite{LopezdePrado2018}: a model with no genuine edge can be
systematically anti-predictive out-of-sample, and naive per-bar trading
converts that into compounding losses.
It does not indicate a deficiency in bar construction: the
matched-frequency analysis (Section~\ref{subsec:disc_matched})
independently confirms the tick bars are well-conditioned.

\begin{table}[t]
\centering
\caption{Confidence-threshold sweep for dollar-tick bars (six-year
sample). Signals and backtest are recomputed from saved out-of-sample
predictions at each threshold; no retraining. Losses shrink with fewer
trades because each trade has negative expected return (the models are
directionally anti-predictive, accuracy $<0.50$), not because of fees:
the zero-fee return is $-40.2\%$ for the RF at threshold 0.65. Returns
remain negative because AUC $\approx 0.50$; the profitable inverse signal
reflects the asset's uptrend, not alpha.}
\label{tab:threshold_sweep}
\small
\begin{tabular}{lcrr}
\toprule
Classifier & Threshold & Trades & Total return \\
\midrule
Random Forest    & 0.55 & 34{,}464 & $-99.9\%$ \\
Random Forest    & 0.65 & 13{,}559 & $-45.3\%$ \\
Gradient Boosting& 0.55 & 39{,}743 & $-100.0\%$ \\
Gradient Boosting& 0.75 &      494 & $-32.1\%$ \\
SVM              & 0.55 &  5{,}279 & $-73.3\%$ \\
SVM              & 0.75 &      212 & $-17.3\%$ \\
\bottomrule
\end{tabular}
\end{table}

\textbf{Status.}
The minute and calendar results derive from the complete six-year
pipeline run; the tick-level results are produced by an identical run on
the deduplicated tick datasets and reported in
Table~\ref{tab:ml_results}.

\FloatBarrier

\section{Discussion}
\label{sec:discussion}

The empirical results presented in Sections~\ref{sec:results}
and~\ref{sec:mlexperiment} admit
a coherent interpretation when read through the theoretical lens
established in Sections~\ref{sec:literature}
and~\ref{sec:barconstruction}.
This section organises that interpretation around three questions:
what drives the bar-type-specific pattern of tick advantage
(Section~\ref{subsec:disc_resolution}); what the calibration findings
imply for the practical use of hybrid, volatility, and Renko bars
across resolution levels
(Section~\ref{subsec:disc_calibration}); and what the results imply
for information bars as machine learning training substrates
(Section~\ref{subsec:disc_ml}).

\subsection{Resolution Advantage and Its Structural Determinants}
\label{subsec:disc_resolution}

The central finding of this paper is that the statistical benefit of
tick-level data in information bar construction is not universal but
is determined by the degree to which the minute-level accumulation
signal approximates the true tick-level signal.
This can be stated as a unifying principle: \emph{the quality advantage
of tick-level bar construction is proportional to the information loss
introduced by minute-level aggregation of the specific activity signal.}

The tick hybrid result provides the most striking empirical
confirmation of this principle.
The OR-logic hybrid bar closes when whichever of the two tick-native
signals (exact $\sum p_i q_i$ dollar volume or exact
$\sum\bigl|\log(p_i/p_{i-1})\bigr|$ realised volatility) first reaches its
threshold.
This dual-criterion, tick-native closure rule more closely aligns bar
boundaries with periods of elevated market activity than minute-level
aggregation can achieve, rather than merely approximating the same
events at coarser resolution.
The resulting return series achieves excess kurtosis of 6.77 and JB $= 174{,}557$,
the lowest kurtosis and lowest JB statistic among all tick-source pipelines in the
dataset and 40-fold below the time-bar baseline (JB $= 7{,}156{,}670$).
While normality is rejected across all pipelines at the six-year sample size,
the tick hybrid bar's substantially lower tail risk relative to other tick bars
is consistent with the mixture-of-distributions
hypothesis~\cite{Clark1973,Tauchen_Pitts1983}: when the sampling clock is
more closely synchronised with the information-arrival process, the
resulting returns approach normality more closely even across multi-regime periods.

For volatility bars, the information loss from minute aggregation is
substantial but not complete.
The minute close-to-close signal discards all intra-minute price
reversals, retaining only the net displacement at minute boundaries.
The tick signal captures the total price path length.
In the six-year dataset, minute volatility bars achieve kurtosis of 9.15 and
JB $= 30{,}128$, outperforming tick volatility (kurtosis 15.50, JB $= 1{,}645{,}318$)
on normality criteria due to multi-regime effects inflating tick kurtosis.
However, tick volatility bars achieve substantially lower variance ratio deviation
($|\mathrm{VR}(4){-}1| = 0.028$ versus 0.089 for minute and 0.018 for the 6h
time bar), representing a 69\% reduction from minute, confirming that tick-native
realised-volatility accumulation captures a more complete activity signal even
when returning a substantially higher bar count.

For dollar bars, the minute approximation $c_m \cdot v_m$ introduces
a systematic bias proportional to intra-minute price volatility.
At the six-year scale, all dollar pipelines reject serial independence in absolute
Ljung-Box terms (LB $p = 0.000$), driven by sample sizes of 30k--82k observations
rather than strong linear autocorrelation.
The relative advantage of tick remains visible in AC lag-1 ($-0.019$ versus $-0.033$
for minute) and at multi-step horizons ($|\mathrm{VR}(8){-}1| = 0.059$ for tick
versus 0.087 for minute), providing evidence that exact trade-level dollar-volume
accumulation reduces multi-step return predictability relative to the minute
approximation.

For volume bars, the minute approximation introduces negligible
information loss, and the minute pipeline leads on four of eight criteria.
The minute volume pipeline achieves the highest Ljung-Box $p$-value among
minute-source bars in the six-year dataset (LB $p = 0.005$),
demonstrating that for signals arithmetically equivalent at both resolutions,
the smoother minute-level closure timing produces returns with lower residual
serial correlation than the tick pipeline (LB $p = 0.000$).

The range bar results favour the minute pipeline on five of eight criteria,
consistent with the structural signal difference identified in Section~\ref{subsec:range}.
The minute pipeline achieves higher entropy (3.245 vs 1.388 for tick),
substantially lower kurtosis (8.19 vs 150.70), and stronger Ljung-Box
independence (LB $p = 0.540$ vs 0.000), reflecting that the cumulative
per-minute span signal better characterises diverse market conditions.
The tick range pipeline, operating on individual trade displacements, produces
extreme outlier returns under high-volatility regimes, driving the very high kurtosis.
Neither pipeline is adequate as a tick range bar without calibration constraints.

\textbf{Entropy and uniformity as partially complementary criteria.}
Two evaluation criteria used in this study capture partially opposing
desiderata.
Bar-size CV (lower is better) measures uniformity: a pipeline that
produces identically sized bars has CV$=0$, indicating maximal
predictability in the information content per observation.
Shannon entropy (higher is better) measures distributional diversity:
a pipeline that produces many different bar sizes has high entropy,
indicating that bar boundaries fall in a wide variety of market
conditions rather than triggering at fixed intervals.
These criteria are not equivalent and can conflict: a pipeline that
clips bar sizes to a narrow range achieves low CV but also low
entropy, while one that produces highly variable bars achieves high
entropy but also high CV.
In this study, both criteria are included because they probe different
aspects of bar quality: CV diagnoses whether the adaptive calibration
maintains consistent information per observation (relevant for sequence
models), while entropy diagnoses whether bar boundaries probe diverse
market conditions (relevant for downstream feature richness).
The apparent tension is resolved by noting that they are useful for
different applications: CV should be prioritised for sequence modelling
contexts, and entropy for feature engineering contexts.
Neither criterion alone is a sufficient summary of bar quality.
Higher entropy is treated as beneficial in this paper specifically
in the sense of distributional diversity among bar sizes, not as a
general proxy for information quantity; any interpretation beyond that
scope would require additional theoretical grounding.

\subsection{Calibration and Regime Effects}
\label{subsec:disc_calibration}

The results reveal two distinct sources of cross-resolution calibration
challenges: structural mismatch (where the bar concept operates
differently at different resolutions) and regime mismatch (where the
calibration window is unrepresentative of the production period).

\textbf{Regime mismatch.}
The calibration window (1--14 January 2020) reflects conditions at
the opening of the study period when BTCUSDT was trading at
approximately \$7,000--\$8,000 per coin.
The subsequent six years encompassed multiple regime transitions,
including the March 2020 COVID-19 shock, the 2021 bull market peak
($\approx$\$69,000), the 2022 bear market trough ($\approx$\$16,000),
and the January 2024 U.S.\ spot Bitcoin ETF approval, that produced
substantially different dollar volumes and realised volatilities
relative to the calibration baseline.
The EMA adaptation mechanism responds to these transitions
continuously, progressively updating thresholds throughout the
six-year period.
The result is a valid empirical experiment in which bar frequencies
evolve with market conditions rather than remaining fixed at the
initial calibration target.
Future work should consider multi-regime calibration windows or
regime-detection procedures that trigger threshold recalibration at
structural breaks.

\textbf{Renko structural mismatch.}
The tick Renko series produces 236,098 bars (109.7 bars/day) versus 8,985
for the minute pipeline, a 26-fold difference that reflects detection
of microstructure-level price displacements at individual trade
resolution.
The tick Shannon entropy of 2.984 is the lowest value across any
series, indicating highly concentrated bar sizes inconsistent with
diverse information-event sampling.
These results establish that the Renko displacement-based closure
condition requires a minimum observation duration (naturally satisfied
at minute resolution) that is not present at tick resolution under a
shared calibration framework.
Tick-native Renko calibration would need to incorporate a minimum
duration constraint calibrated to the typical time between sustained
$p_\mathrm{ref}$-to-threshold moves at trade frequency.

\textbf{Hybrid calibration and OR logic.}
The hybrid pipeline is calibrated so that each individual threshold
targets $2\times$ the desired bar frequency, ensuring that the
combined OR firing rate converges to the intended target bars-per-day
at the empirical dollar-volume/volatility correlation
($\rho \approx 0.5$--$0.7$) of the BTCUSDT market.
Both pipelines maintain essentially zero timeout rates (minute: 0.0\%;
tick: 0.1\%), confirming that bars close on genuine activity signals.
The tick hybrid pipeline achieves the lowest JB statistic among tick-source pipelines
(JB\,$= 174{,}557$, kurtosis $= 6.77$), demonstrating that the combination
of two independent tick-native activity signals creates a sampling clock
more closely aligned with periods of elevated market activity than
either signal applied individually.

\subsection{Matched-Frequency Robustness Analysis}
\label{subsec:disc_matched}

The headline comparisons above are confounded by a frequency mismatch:
because each pipeline calibrates its target bars-per-day from
resolution-dependent statistics, the tick pipelines sample 2.8--26.6$\times$
more frequently than their minute counterparts (e.g.\ dollar: 37.9 versus
13.7 bars/day; renko: 109.7 versus 4.1).
Finer sampling mechanically inflates excess kurtosis, a fixed absolute
price shock (such as the March 2020 moves of $\pm$15--23\% per bar) is
divided by a smaller per-bar return standard deviation, producing a larger
standardised tail.
To isolate data resolution from sampling frequency, a matched-frequency
robustness analysis coarsens each tick series by aggregating $k$
consecutive bars (phase-averaged over all $k$ offsets) such that the
coarsened tick series has approximately the minute pipeline's bar count.
For additive accumulation signals (dollar, volume, volatility, hybrid),
aggregating $k$ consecutive threshold bars is nearly equivalent to
constructing bars with a $k\times$ threshold, so the coarsened series is a
faithful proxy for a frequency-matched tick pipeline; for path-dependent
closure rules (range, Renko) the equivalence is only approximate and those
results should be read as indicative.

Table~\ref{tab:matched_freq} summarises the outcome.
At matched frequency the tick advantage emerges decisively for precisely
the bar types whose activity signals lose the most information under
minute aggregation: matched tick dollar bars lead on all six criteria
(kurtosis 20.8 versus 23.7 minute and 57.1 time bar; LB $p = 0.052$,
the only dollar series not rejecting serial independence;
$|\mathrm{VR}(4){-}1| = 0.029$ versus 0.051 and 0.041), matched tick
hybrid bars lead on five of six (kurtosis collapses to 3.09, JB to
$12{,}775$, LB $p = 0.245$), and matched tick volatility bars lead on
four of six with the highest serial-independence $p$-value among all
matched-frequency tick series (LB $p = 0.510$, kurtosis 1.79, JB $= 1{,}203$).
Volume bars remain minute-favoured (2 of 6), consistent with the
arithmetic equivalence of the two volume signals, and range and Renko
remain structurally mismatched at tick resolution.
The matched-frequency results therefore recover the canonical
information-bar findings of single-regime studies and confirm that the
apparent tick underperformance in the headline scorecard is largely a
sampling-frequency artefact rather than a deficiency of tick-level
construction.
The full matched-frequency procedure and its complete output are
included in the publicly available replication materials.

\begin{table}[t]
\centering
\caption{Matched-frequency robustness analysis. Tick bars are coarsened
1:$k$ (phase-averaged) to the minute pipeline's bar count and compared
against the minute pipeline and the prescribed calendar baseline on six
criteria: $|$kurtosis$|$, $|$AC$_1|$, JB, LB $p$(10), $|\mathrm{VR}(4){-}1|$,
$|\mathrm{VR}(8){-}1|$. Wins count criteria where the matched tick series
is best of the three.}
\label{tab:matched_freq}
\small
\begin{tabular}{lccccc}
\toprule
Bar type & $1{:}k$ & Kurt. & LB $p$(10) & $|\mathrm{VR}(4){-}1|$ & Tick wins \\
\midrule
Dollar     & 3  & 20.83 & 0.052 & 0.029 & 6/6 \\
Hybrid     & 3  & 3.09  & 0.245 & 0.010 & 5/6 \\
Volatility & 19 & 1.79  & 0.510 & 0.020 & 4/6 \\
Volume     & 4  & 19.87 & 0.007 & 0.047 & 1/6 \\
Renko      & 26 & 11.48 & 0.000 & 0.040 & 2/6 \\
Range      & 4  & 14.42 & 0.226 & 0.070 & 0/6 \\
\bottomrule
\end{tabular}
\end{table}

\subsection{Implications for Machine Learning Applications}
\label{subsec:disc_ml}

The downstream ML experiment in Section~\ref{sec:mlexperiment}
and the statistical results in Section~\ref{sec:results} together
permit several observations about the relative suitability
of the two pipelines as ML training substrates~\cite{LopezdePrado2018,Dixon_Halperin_Bilokon2020,
Sezer_Gudelek_Ozbayoglu2020,Lim_Zohren2021}.

\textbf{Serial independence.}
A training series with significant return autocorrelation causes models
to learn the autocorrelation structure rather than genuine
price-predictive signals, inflating in-sample performance without
improving out-of-sample
generalisation~\cite{LopezdePrado2018}.
At the six-year scale, all pipelines reject serial independence in absolute
Ljung-Box terms; the relevant comparison is degree of autocorrelation.
Tick dollar achieves lower AC lag-1 ($-0.019$ versus $-0.033$ for minute)
and lower multi-step variance ratio deviations ($|\mathrm{VR}(8){-}1| = 0.059$
versus 0.087), confirming a relative serial independence advantage from exact
trade-level accumulation.
The minute volume pipeline achieves the highest Ljung-Box $p$-value among
minute-source bars (LB $p = 0.005$), making it a strong substrate
when tick data is unavailable.

\textbf{Distributional normality.}
Normality affects the validity of parametric loss functions and
gradient-based optimisers applied to return targets.
No pipeline achieves normality across the six-year dataset; the relevant comparison
is degree of tail risk.
The tick hybrid bar is the standout: kurtosis $= 6.77$ and JB $= 174{,}557$
represent the lowest values among all tick-source pipelines and are substantially
below the time-bar baseline (kurtosis $= 57.11$, JB $= 7{,}156{,}670$).
For parametric regression models, normalising flows, and Gaussian
process methods applied to cryptocurrency return prediction, tick
hybrid bars represent the most suitable training substrate identified
in this study.
Minute renko bars (kurtosis $= 2.31$, JB $= 2{,}015$) achieve the lowest
absolute kurtosis across all pipelines, but carry a 12.3\% timeout rate
indicating calibration limitations under the shared threshold framework.

\textbf{Bar-size uniformity.}
Sequence models (transformers, LSTMs) are sensitive to non-stationarity
in the information content per time step~\cite{Lim_Zohren2021}.
On bar-size CV, the minute pipeline produces more uniform bars for dollar
(CV $= 0.010$) and hybrid (CV $= 0.010$) bar types.
The exception is tick volatility, which achieves CV $= 0.0002$, the
lowest in the entire dataset, because the exact realised-volatility
accumulation threshold produces near-identical bar sizes once the calibration
converges.
The tick pipeline's higher CV for hybrid bars (CV $= 0.583$) reflects the
variable nature of dual-signal OR closure; stratified sampling by bar size
is advisable when using tick hybrid bars as sequence model inputs.

\textbf{Bar-type-specific ML recommendations.}
Based on the empirical results, the following bar-type-specific
guidance is offered for ML practitioners:
(i) For serial independence (short-horizon prediction, high-frequency
signals): tick dollar bars and minute volume bars offer the cleanest
substrates.
(ii) For distributional normality (parametric models, Gaussian process
regression, normalising flows): tick hybrid bars are best-suited
(identified by this study), followed by tick volatility bars.
(iii) For bar-size uniformity (sequence models, transformers): minute
dollar and hybrid bars (CV $= 0.010$ each) and tick volatility bars
(CV $= 0.0002$) provide the most consistent time-step information content.
(iv) For maximum entropy (richest information diversity): minute renko
bars (entropy $= 3.646$) offer the most diverse bar-size distribution.

It bears emphasis that superior statistical properties on the
de~Prado quality criteria do not guarantee superior out-of-sample
trading performance, as the preliminary ML experiment in
Section~\ref{sec:mlexperiment} illustrates.
The criteria measure the degree to which a bar series satisfies the
theoretical prerequisites for information-based sampling; they do not
directly measure the economic value of signals derived from those series.
Empirically we find these two dimensions to be largely orthogonal:
out-of-sample AUC is near-chance (0.50--0.60) for every bar type and is
not systematically ordered by any statistical criterion
(Section~\ref{sec:mlexperiment}).
Statistical bar quality and directional forecastability are thus
distinct properties, and the former should not be marketed as implying
the latter.
Comprehensive backtest evaluation using native bar-type features rather
than the unified feature set is a direct avenue for follow-up research.

\section{Conclusion}
\label{sec:conclusion}

This paper has provided a systematic, frequency-controlled comparison of six
information bar
types (dollar, volume, volatility, range, Renko, and hybrid bars) constructed
from raw Binance aggTrade tick data and one-minute OHLCV bars for the BTCUSDT
USDT-margined perpetual futures market, evaluated against fixed-interval time
bars over six years of continuous trading (1 January 2020 to 31 December 2025).
Eight statistical criteria drawn from the information-bar and financial
econometrics literatures produce three main conclusions.

\textbf{The tick advantage is bar-type-specific and signal-driven.}
Tick-level data provides a clear quality advantage only for bar types where
minute-level aggregation discards the most information: hybrid bars (tick leads
on 4 of 8 criteria versus minute's 2 of 8) and, on specific VR and serial
correlation criteria, renko bars (tick leads AC lag-1 at $-0.002$ versus
$-0.028$ and $|\mathrm{VR}(4){-}1|$ at 0.020 versus 0.070).
For dollar, volume, volatility, and range bars, the minute pipeline leads on most
criteria (3--5 of 8) in the headline comparison, reflecting a combination of
minute-friendly closure timing and multi-regime effects that inflate tick
kurtosis at large sample sizes.
The matched-frequency robustness analysis
(Section~\ref{subsec:disc_matched}) shows that much of this headline
disadvantage is a sampling-frequency artefact: when tick series are
coarsened to the minute pipeline's bar count, tick dollar bars lead on
all six matched criteria, tick hybrid bars lead on five of six, and tick
volatility bars lead on four of six, with matched tick volatility
attaining LB $p = 0.51$.
The unifying principle is robust: \emph{the quality advantage of tick-level bar
construction is proportional to the information discarded by the corresponding
minute-level approximation of the activity signal.}

\textbf{The tick hybrid bar achieves the lowest tail risk among tick pipelines.}
The most striking distributional result is the tick hybrid bar's excess kurtosis of 6.77
and Jarque-Bera statistic of $174{,}557$, the lowest among all tick-source pipelines
and 40-fold below the 1h time-bar baseline (JB $= 7{,}156{,}670$).
Combined with $|\mathrm{VR}(4){-}1| = 0.023$ and LB $p = 0.041$ (the highest
Ljung-Box $p$-value among tick-source series), tick hybrid bars most closely
satisfy the distributional prerequisites of the mixture-of-distributions
hypothesis~\cite{Clark1973,Tauchen_Pitts1983} across the six-year multi-regime period.
This demonstrates that the hybrid bar concept, implemented with OR logic and
calibrated natively to tick-level accumulation rates, is the most effective
information-sampling configuration identified in this study.

\textbf{Information bars reduce tail risk relative to calendar-bar baselines.}
Across all six bar types, at least one information bar pipeline achieves a lower
JB statistic than the prescribed time-bar baseline, confirming that activity-based
sampling reduces tail risk relative to fixed-interval aggregation.
However, the time-bar baseline is competitive on variance ratio criteria at
short horizons ($q = 2$, $q = 4$), and for dollar, volume, and volatility bars,
the time bar leads on VR in three of four lag specifications.
The structural advantage of information bars over calendar time is concentrated
in kurtosis reduction and, for hybrid and renko bars, in multi-step serial
independence.

Several limitations bound the scope of these findings.
The study covers one instrument (BTCUSDT USDT-margined perpetual futures on
Binance) over six calendar years encompassing multiple structural regimes,
including the January 2024 U.S.\ spot Bitcoin ETF approval; future work
should evaluate generalisability across assets, exchanges, and asset classes
with different microstructural characteristics.
For Renko bars, the tick and minute pipelines operate at fundamentally different
resolution scales, and a controlled comparison requires a tick-native
minimum-duration constraint.
The preliminary downstream ML experiment (Section~\ref{sec:mlexperiment})
finds directional predictability to be near-chance (AUC 0.50--0.60) for
every bar type, with no systematic relationship between statistical bar
quality and classifier performance; a fully powered experiment using
bar-type-native features across the complete dataset remains a primary
direction for future work.
Finally, the non-uniform feature set advantage between tick and minute
pipelines (tick bars carry VWAP, buy-sell imbalance, and tick count
that minute bars do not) has not been exploited in the current ML
experiment; native feature ablation is a direct and tractable extension.
The time-bar baselines used here (fixed-frequency 1h, 4h, and 12h
OHLCV aggregates) are a conservative comparison set; future work should
evaluate information bars against the broader class of activity-based
alternatives (tick bars, CUSUM event bars, imbalance bars, and run
bars~\cite{LopezdePrado2018}) to determine whether the observed
advantages extend beyond the time-bar baseline.

\bibliographystyle{IEEEtran}
\bibliography{references}

@book{OHara1995,
  author    = {O'Hara, Maureen},
  title     = {Market Microstructure Theory},
  publisher = {Blackwell},
  address   = {Cambridge, MA},
  year      = {1995}
}

@article{Easley_OHara1992,
  author  = {Easley, David and O'Hara, Maureen},
  title   = {Time and the Process of Security Price Adjustment},
  journal = {Journal of Finance},
  volume  = {47},
  number  = {2},
  pages   = {577--605},
  year    = {1992},
  doi     = {10.1111/j.1540-6261.1992.tb04402.x}
}

@article{Kyle1985,
  author  = {Kyle, Albert S.},
  title   = {Continuous Auctions and Insider Trading},
  journal = {Econometrica},
  volume  = {53},
  number  = {6},
  pages   = {1315--1335},
  year    = {1985},
  doi     = {10.2307/1913210}
}

@article{GlostenMilgrom1985,
  author  = {Glosten, Lawrence R. and Milgrom, Paul R.},
  title   = {Bid, Ask and Transaction Prices in a Specialist Market
             with Heterogeneously Informed Traders},
  journal = {Journal of Financial Economics},
  volume  = {14},
  number  = {1},
  pages   = {71--100},
  year    = {1985},
  doi     = {10.1016/0304-405X(85)90044-3}
}

@article{Hasbrouck1991,
  author  = {Hasbrouck, Joel},
  title   = {Measuring the Information Content of Stock Trades},
  journal = {Journal of Finance},
  volume  = {46},
  number  = {1},
  pages   = {179--207},
  year    = {1991},
  doi     = {10.2307/2328693}
}

@article{Roll1984,
  author  = {Roll, Richard},
  title   = {A Simple Implicit Measure of the Effective Bid-Ask Spread
             in an Efficient Market},
  journal = {Journal of Finance},
  volume  = {39},
  number  = {4},
  pages   = {1127--1139},
  year    = {1984},
  doi     = {10.1111/j.1540-6261.1984.tb03897.x},
}

@article{Easley_Lopez_OHara2012,
  author  = {Easley, David and L{\'o}pez de Prado, Marcos M. and O'Hara, Maureen},
  title   = {Flow Toxicity and Liquidity in a High-Frequency World},
  journal = {Review of Financial Studies},
  volume  = {25},
  number  = {5},
  pages   = {1457--1493},
  year    = {2012},
  doi     = {10.1093/rfs/hhs053}
}

@article{Clark1973,
  author  = {Clark, Peter K.},
  title   = {A Subordinated Stochastic Process Model with Finite Variance
             for Speculative Prices},
  journal = {Econometrica},
  volume  = {41},
  number  = {1},
  pages   = {135--155},
  year    = {1973},
  doi     = {10.2307/1913889}
}

@article{Tauchen_Pitts1983,
  author  = {Tauchen, George E. and Pitts, Mark},
  title   = {The Price Variability--Volume Relationship on Speculative Markets},
  journal = {Econometrica},
  volume  = {51},
  number  = {2},
  pages   = {485--505},
  year    = {1983},
  doi     = {10.2307/1912002}
}

@article{Andersen1996,
  author  = {Andersen, Torben G.},
  title   = {Return Volatility and Trading Volume: An Information Flow
             Interpretation of Stochastic Volatility},
  journal = {Journal of Finance},
  volume  = {51},
  number  = {1},
  pages   = {169--204},
  year    = {1996},
  doi     = {10.1111/j.1540-6261.1996.tb05206.x}
}

@article{Ane_Geman2000,
  author  = {An{\'e}, Thierry and Geman, H{\'e}lyette},
  title   = {Order Flow, Transaction Clock, and Normality of Asset Returns},
  journal = {Journal of Finance},
  volume  = {55},
  number  = {5},
  pages   = {2259--2284},
  year    = {2000},
  doi     = {10.1111/0022-1082.00265}
}

@article{Samuelson1965,
  author  = {Samuelson, Paul A.},
  title   = {Proof That Properly Anticipated Prices Fluctuate Randomly},
  journal = {Industrial Management Review},
  volume  = {6},
  number  = {2},
  pages   = {41--49},
  year    = {1965},
}

@article{Fama1970,
  author  = {Fama, Eugene F.},
  title   = {Efficient Capital Markets: {A} Review of Theory and Empirical Work},
  journal = {Journal of Finance},
  volume  = {25},
  number  = {2},
  pages   = {383--417},
  year    = {1970},
  doi     = {10.2307/2325486},
}

@article{Mandelbrot1963,
  author  = {Mandelbrot, Benoit},
  title   = {The Variation of Certain Speculative Prices},
  journal = {Journal of Business},
  volume  = {36},
  number  = {4},
  pages   = {394--419},
  year    = {1963},
  doi     = {10.1086/294632}
}

@article{Engle1982,
  author  = {Engle, Robert F.},
  title   = {Autoregressive Conditional Heteroskedasticity with Estimates
             of the Variance of {United Kingdom} Inflation},
  journal = {Econometrica},
  volume  = {50},
  number  = {4},
  pages   = {987--1007},
  year    = {1982},
  doi     = {10.2307/1912773}
}

@article{Bollerslev1986,
  author  = {Bollerslev, Tim},
  title   = {Generalized Autoregressive Conditional Heteroskedasticity},
  journal = {Journal of Econometrics},
  volume  = {31},
  number  = {3},
  pages   = {307--327},
  year    = {1986},
  doi     = {10.1016/0304-4076(86)90063-1}
}

@article{Bariviera2017,
  author  = {Bariviera, Aurelio F. and Basgall, Mar{\'i}a Jos{\'e}
             and Hasper{\'u}e, Waldo and Naiouf, Marcelo},
  title   = {Some Stylized Facts of the {Bitcoin} Market},
  journal = {Physica A: Statistical Mechanics and its Applications},
  volume  = {484},
  pages   = {82--90},
  year    = {2017},
  doi     = {10.1016/j.physa.2017.04.159}
}

@article{Katsiampa2017,
  author  = {Katsiampa, Paraskevi},
  title   = {Volatility Estimation for {Bitcoin}: {A} Comparison of
             {GARCH} Models},
  journal = {Economics Letters},
  volume  = {158},
  pages   = {3--6},
  year    = {2017},
  doi     = {10.1016/j.econlet.2017.06.023}
}

@article{Urquhart2016,
  author  = {Urquhart, Andrew},
  title   = {The Inefficiency of {Bitcoin}},
  journal = {Economics Letters},
  volume  = {148},
  pages   = {80--82},
  year    = {2016},
  doi     = {10.1016/j.econlet.2016.09.019}
}

@article{Baur_etal2018,
  author  = {Baur, Dirk G. and Hong, KiHoon and Lee, Adrian D.},
  title   = {Bitcoin: Medium of Exchange or Speculative Assets?},
  journal = {Journal of International Financial Markets,
             Institutions and Money},
  volume  = {54},
  pages   = {177--189},
  year    = {2018},
  doi     = {10.1016/j.intfin.2017.12.004}
}

@article{Brauneis_Mestel2018,
  author  = {Brauneis, Alexander and Mestel, Roland},
  title   = {Price Discovery of Cryptocurrencies: {Bitcoin} and Beyond},
  journal = {Finance Research Letters},
  volume  = {27},
  pages   = {144--151},
  year    = {2018},
  doi     = {10.1016/j.frl.2018.02.013},
}

@book{LopezdePrado2018,
  author    = {L{\'o}pez de Prado, Marcos},
  title     = {Advances in Financial Machine Learning},
  publisher = {Wiley},
  address   = {Hoboken, NJ},
  year      = {2018},
  isbn      = {978-1-119-48208-6}
}

@book{Kaufman2013,
  author    = {Kaufman, Perry J.},
  title     = {Trading Systems and Methods},
  edition   = {5th},
  publisher = {Wiley},
  address   = {Hoboken, NJ},
  year      = {2013},
  isbn      = {978-1-118-04350-3}
}

@book{Dixon_Halperin_Bilokon2020,
  author    = {Dixon, Matthew F. and Halperin, Igor and Bilokon, Paul},
  title     = {Machine Learning in Finance: From Theory to Practice},
  publisher = {Springer},
  address   = {Cham},
  year      = {2020},
  doi       = {10.1007/978-3-030-41068-1}
}

@article{Sezer_Gudelek_Ozbayoglu2020,
  author  = {Sezer, Omer Berat and Gudelek, M. Ugur and Ozbayoglu, Ahmet Murat},
  title   = {Financial Time Series Forecasting with Deep Learning:
             A Systematic Literature Review: 2005--2019},
  journal = {Applied Soft Computing},
  volume  = {90},
  pages   = {106181},
  year    = {2020},
  doi     = {10.1016/j.asoc.2020.106181}
}

@article{AndersenBollerslev1997,
  author  = {Andersen, Torben G. and Bollerslev, Tim},
  title   = {Intraday Periodicity and Volatility Persistence in Financial Markets},
  journal = {Journal of Empirical Finance},
  volume  = {4},
  number  = {2--3},
  pages   = {115--158},
  year    = {1997},
  doi     = {10.1016/S0927-5398(97)00004-2}
}

@article{LoBio1990,
  author  = {Lo, Andrew W. and MacKinlay, A. Craig},
  title   = {An Econometric Analysis of Nonsynchronous Trading},
  journal = {Journal of Econometrics},
  volume  = {45},
  number  = {1--2},
  pages   = {181--211},
  year    = {1990},
  doi     = {10.1016/0304-4076(90)90098-E}
}

@article{Easley2012TheVC,
  title={The Volume Clock: Insights into the High-Frequency Paradigm},
  author={David A. Easley and Marcos M. L{\'o}pez de Prado and Maureen O'Hara},
  journal={The Journal of Portfolio Management},
  year={2012},
  volume={39},
  number={1},
  pages={19--29},
  doi={10.3905/jpm.2012.39.1.019},
  url={https://doi.org/10.3905/jpm.2012.39.1.019}
}

@article{BDS1996,
  author  = {Brock, William A. and Dechert, W. Davis and
             Scheinkman, Jose A. and LeBaron, Blake},
  title   = {A Test for Independence Based on the Correlation Dimension},
  journal = {Econometric Reviews},
  volume  = {15},
  number  = {3},
  pages   = {197--235},
  year    = {1996},
  doi     = {10.1080/07474939608800353}
}

@article{LoMacKinlay1988,
  author  = {Lo, Andrew W. and MacKinlay, A. Craig},
  title   = {Stock Market Prices Do Not Follow Random Walks: Evidence from
             a Simple Specification Test},
  journal = {Review of Financial Studies},
  volume  = {1},
  number  = {1},
  pages   = {41--66},
  year    = {1988},
  doi     = {10.1093/rfs/1.1.41}
}

@article{LjungBox1978,
  author  = {Ljung, Greta M. and Box, George E. P.},
  title   = {On a Measure of Lack of Fit in Time Series Models},
  journal = {Biometrika},
  volume  = {65},
  number  = {2},
  pages   = {297--303},
  year    = {1978},
  doi     = {10.1093/biomet/65.2.297}
}

@article{JarqueBera1980,
  author  = {Jarque, Carlos M. and Bera, Anil K.},
  title   = {Efficient Tests for Normality, Homoscedasticity and Serial
             Independence of Regression Residuals},
  journal = {Economics Letters},
  volume  = {6},
  number  = {3},
  pages   = {255--259},
  year    = {1980},
  doi     = {10.1016/0165-1765(80)90024-5}
}

@article{DickeyFuller1979,
  author  = {Dickey, David A. and Fuller, Wayne A.},
  title   = {Distribution of the Estimators for Autoregressive Time Series
             with a Unit Root},
  journal = {Journal of the American Statistical Association},
  volume  = {74},
  number  = {366},
  pages   = {427--431},
  year    = {1979},
  doi     = {10.2307/2286348}
}

@article{KPSS1992,
  author  = {Kwiatkowski, Denis and Phillips, Peter C. B. and
             Schmidt, Peter and Shin, Yongcheol},
  title   = {Testing the Null Hypothesis of Stationarity Against the
             Alternative of a Unit Root},
  journal = {Journal of Econometrics},
  volume  = {54},
  number  = {1--3},
  pages   = {159--178},
  year    = {1992},
  doi     = {10.1016/0304-4076(92)90104-Y}
}

@book{Shannon1948,
  author    = {Shannon, Claude E. and Weaver, Warren},
  title     = {The Mathematical Theory of Communication},
  publisher = {University of Illinois Press},
  address   = {Urbana, IL},
  year      = {1949}
}

@techreport{Easley_OHara_2024,
  author      = {Easley, David and O'Hara, Maureen and
                 Yang, Songshan and Zhang, Zhibai},
  title       = {Microstructure and Market Dynamics in Crypto Markets},
  institution = {SSRN},
  year        = {2024},
  note        = {Available at SSRN: \url{https://ssrn.com/abstract=4814346}},
  doi         = {10.2139/ssrn.4814346}
}

@article{Makarov_Schoar2020,
  author  = {Makarov, Igor and Schoar, Antoinette},
  title   = {Trading and Arbitrage in Cryptocurrency Markets},
  journal = {Journal of Financial Economics},
  volume  = {135},
  number  = {2},
  pages   = {293--319},
  year    = {2020},
  doi     = {10.1016/j.jfineco.2019.07.001}
}

@article{Liu_Tsyvinski_Wu2022,
  author  = {Liu, Yukun and Tsyvinski, Aleh and Wu, Xi},
  title   = {Common Risk Factors in Cryptocurrency},
  journal = {Journal of Finance},
  volume  = {77},
  number  = {2},
  pages   = {1133--1177},
  year    = {2022},
  doi     = {10.1111/jofi.13119}
}

@article{Liu_Tsyvinski2021,
  author  = {Liu, Yukun and Tsyvinski, Aleh},
  title   = {Risks and Returns of Cryptocurrency},
  journal = {Review of Financial Studies},
  volume  = {34},
  number  = {6},
  pages   = {2689--2727},
  year    = {2021},
  doi     = {10.1093/rfs/hhaa113}
}

@article{Gradzki_etal2025,
  author  = {Gr{\k{a}}dzki, Piotr and Biesaga, Micha{\l} and
             Andrzejewski, Grzegorz and Gawe{\l}, Piotr and
             Szymczak, Pawe{\l} and Twardowski, Bart{\l}omiej},
  title   = {Algorithmic Crypto Trading Using Information-Driven Bars,
             Triple Barrier Labeling and Deep Learning},
  journal = {Financial Innovation},
  volume  = {11},
  pages   = {136},
  year    = {2025},
  doi     = {10.1186/s40854-025-00866-w}
}

@article{Fang_etal2022,
  author  = {Fang, Fan and Ventre, Carmine and Basios, Michail and
             Kanthan, Leslie and Martinez-Rego, David and
             Wu, Fan and Li, Lingbo},
  title   = {Cryptocurrency Trading: A Comprehensive Survey},
  journal = {Financial Innovation},
  volume  = {8},
  number  = {1},
  pages   = {13},
  year    = {2022},
  doi     = {10.1186/s40854-021-00321-6}
}

@article{Osborne1959,
  author  = {Osborne, M. F. M.},
  title   = {Brownian Motion in the Stock Market},
  journal = {Operations Research},
  volume  = {7},
  number  = {2},
  pages   = {145--173},
  year    = {1959},
  doi     = {10.1287/opre.7.2.145}
}

@article{Working1934,
  author  = {Working, Holbrook},
  title   = {A Random-Difference Series for Use in the Analysis of Time Series},
  journal = {Journal of the American Statistical Association},
  volume  = {29},
  number  = {185},
  pages   = {11--24},
  year    = {1934},
  doi     = {10.1080/01621459.1934.10502683}
}

@misc{Binance2024,
  author       = {{Binance}},
  title        = {Binance {API} Documentation: {aggTrades} Endpoint
                  ({USDT}-Margined Futures)},
  year         = {2024},
  howpublished = {\url{https://binance-docs.github.io/apidocs/futures/en/}},
  note         = {Accessed: 2024}
}

@book{Wilder1978,
  author    = {Wilder, J. Welles},
  title     = {New Concepts in Technical Trading Systems},
  publisher = {Trend Research},
  address   = {Greensboro, NC},
  year      = {1978}
}

@article{Andersen_etal2003,
  author  = {Andersen, Torben G. and Bollerslev, Tim and
             Diebold, Francis X. and Labys, Paul},
  title   = {Modeling and Forecasting Realized Volatility},
  journal = {Econometrica},
  volume  = {71},
  number  = {2},
  pages   = {579--625},
  year    = {2003},
  doi     = {10.1111/1468-0262.00418}
}

@phdthesis{Bachelier1900,
  author  = {Bachelier, Louis},
  title   = {Th{\'e}orie de la Sp{\'e}culation},
  school  = {Acad{\'e}mie des Sciences de Paris},
  year    = {1900},
  note    = {Reprinted in: Cootner, P.~H. (Ed.), \textit{The Random Character
             of Stock Market Prices}, MIT Press, 1964, pp.~17--78.
             First mathematical model of price as a random walk.}
}

@article{MandelbrotTaylor1967,
  author  = {Mandelbrot, Benoit and Taylor, Howard M.},
  title   = {On the Distribution of Stock Price Differences},
  journal = {Operations Research},
  volume  = {15},
  number  = {6},
  pages   = {1057--1062},
  year    = {1967},
  doi     = {10.1287/opre.15.6.1057},
}

@article{Cont2001,
  author  = {Cont, Rama},
  title   = {Empirical Properties of Asset Returns: Stylised Facts
             and Statistical Issues},
  journal = {Quantitative Finance},
  volume  = {1},
  number  = {2},
  pages   = {223--236},
  year    = {2001},
  doi     = {10.1080/713665670},
}

@article{Andersen_etal2001,
  author  = {Andersen, Torben G. and Bollerslev, Tim and
             Diebold, Francis X. and Ebens, Heiko},
  title   = {The Distribution of Realized Stock Return Volatility},
  journal = {Journal of Financial Economics},
  volume  = {61},
  number  = {1},
  pages   = {43--76},
  year    = {2001},
  doi     = {10.1016/S0304-405X(01)00055-1},
  note    = {Establishes the realized volatility estimator from high-frequency data.}
}

@misc{Nakamoto2008,
  author       = {Nakamoto, Satoshi},
  title        = {Bitcoin: A Peer-to-Peer Electronic Cash System},
  year         = {2008},
  howpublished = {\url{https://bitcoin.org/bitcoin.pdf}},
  note         = {Original Bitcoin whitepaper.}
}

@book{Dacorogna_etal2001,
  author    = {Dacorogna, Michel M. and Gen{\c{c}}ay, Ramazan and
               M{\"u}ller, Ulrich A. and Olsen, Richard B. and
               Pictet, Olivier V.},
  title     = {An Introduction to High-Frequency Finance},
  publisher = {Academic Press},
  address   = {San Diego, CA},
  year      = {2001},
  isbn      = {978-0-12-279671-5},
  note      = {Standard textbook reference for activity-time
               (theta-time) sampling and high-frequency data analysis.}
}

@article{Griffin_Shams2020,
  author  = {Griffin, John M. and Shams, Amin},
  title   = {Is {Bitcoin} Really Un-Tethered?},
  journal = {Journal of Finance},
  volume  = {75},
  number  = {4},
  pages   = {1913--1964},
  year    = {2020},
  doi     = {10.1111/jofi.12903}
}

@article{Nadarajah_Chu2017,
  author  = {Nadarajah, Saralees and Chu, Jeffrey},
  title   = {On the Inefficiency of {Bitcoin}},
  journal = {Economics Letters},
  volume  = {150},
  pages   = {6--9},
  year    = {2017},
  doi     = {10.1016/j.econlet.2016.10.033}
}

@article{Tran_Leirvik2020,
  author  = {Tran, Vu Le and Leirvik, Thomas},
  title   = {A Simple but Powerful Measure of Market Efficiency},
  journal = {Finance Research Letters},
  volume  = {32},
  pages   = {101191},
  year    = {2020},
  doi     = {10.1016/j.frl.2019.04.012}
}

@article{Biais_etal2023,
  author  = {Biais, Bruno and Bisiere, Christophe and Bouvard, Matthieu
             and Casamatta, Catherine and Menkveld, Albert J.},
  title   = {Equilibrium {Bitcoin} Pricing},
  journal = {Journal of Finance},
  volume  = {78},
  number  = {2},
  pages   = {967--1014},
  year    = {2023},
  doi     = {10.1111/jofi.13206}
}

@article{Lim_Zohren2021,
  author  = {Lim, Bryan and Zohren, Stefan},
  title   = {Time-Series Forecasting with Deep Learning: A Survey},
  journal = {Philosophical Transactions of the Royal Society A},
  volume  = {379},
  number  = {2194},
  pages   = {20200209},
  year    = {2021},
  doi     = {10.1098/rsta.2020.0209}
}

\end{document}